\documentclass[10pt,conference]{IEEEtran}
\IEEEoverridecommandlockouts
\usepackage{amsmath,amssymb,amsfonts}
\usepackage{algorithmic}
\usepackage{graphicx}
\usepackage{gensymb}
\usepackage{caption}
\usepackage{subcaption}
\usepackage{adjustbox}
\usepackage{textcomp}
\usepackage{xcolor}
\usepackage{xfrac}
\usepackage{comment}
\usepackage[siunitx]{circuitikz}
\def\BibTeX{{\rm B\kern-.05em{\sc i\kern-.025em b}\kern-.08em
    T\kern-.1667em\lower.7ex\hbox{E}\kern-.125emX}}

\usepackage{multirow}
\usepackage{lipsum}
\begin{document}

\title{\color{black}Analysis and Demonstration of a Current-Stiff, Isolated Capacitive Bridge Transformer\color{black}\\
\thanks{This work was supported in part by USA National Science Foundation Grant No. 2328137. \\ This work has been submitted to the IEEE for possible
publication. Copyright may be transferred without notice, after which this version may no longer be accessible.}
}

\author{
\IEEEauthorblockN{Patrick Nowakowski \color{black}and Cameron Little\color{black}, \textit{Student Members, IEEE}, }
\IEEEauthorblockN{Mahima Gupta and Daniel C. Ludois, \textit{Senior Members, IEEE}}
\IEEEauthorblockA{\textit{Dept. of Electrical and Computer Engineering} \\
\textit{Wisconsin Electric Machines and Power Electronics Consortium (WEMPEC)} \\
\textit{University of Wisconsin-Madison}\\
Madison, WI 53706, USA \\
Corresponding email: ludois@wisc.edu}
}

\maketitle

\begin{abstract}
The traditional Voltage Source Inverter (VSI) based Dual Active Bridge (DAB) utilizes a transformer at its center to provide galvanic isolation and a turns ratio for step up/down functionality.  However, the transformer is often the limiting factor of the DAB, especially as semiconductor switching speeds improve, given steel core losses, winding losses, etc. This paper presents a new DAB architecture based on Current Source Inverters (CSIs) coupled through, and isolated by, an AC capacitive bridge, yielding a system called a Capacitive Bridge Transformer (CBT). The CBT architecture eliminates all magnetic components in the AC portion of the isolated DAB circuit, reducing frequency scaling barriers normally present in DABs. \color{black} The paper introduces the CBT from the perspective of duality and then develops the DC-DC power flow equations, the internal apparent power equations, and effective turns ratio. The analytical models are verified by steady state and detailed time domain measurements on a 2 kW experimental prototype. \color{black}
\end{abstract}

\begin{IEEEkeywords}
DC-DC, Solid State Transformer, Current Source Inverters, Capacitive Bridge
\end{IEEEkeywords}

\section{Introduction}

At the core of most solid-state transformer (SST) technology lies a high-power DC-DC converter, typically the dual active bridge (DAB) \cite{Huber2016Solid-StateConcepts}.  The most common DAB configuration consists of two voltage source inverters (VSIs) whose AC terminals are joined by a transformer. The VSIs enable higher than the line frequency operation, thereby shrinking the transformer.  Since its formal introduction in \cite{Doncker1991AApplications}, the VSI-based DAB has been one of the most researched power electronic architectures of the past 35 years and has been successfully applied in numerous applications. However, it is not without its challenges going forward. The capabilities of wide-bandgap (WBG) semiconductor devices are progressing faster than those of passive components, especially transformers, leaving these semiconductors underutilized within DABs.  Focusing on the transformer as the bottleneck for DAB progression, specific shortcomings include:
\begin{enumerate}
  \item Losses – Transformer core materials limit the maximum operating frequency and reactive circulating power increases winding conduction losses in high-power applications \cite{Guillod2020Medium-frequencyAnalysis, Leibl2017DesignApplications, Soltau2014IronConverter}.
  \item Design – Cross coupling of transformer leakage and magnetizing inductance, i.e. these cannot be chosen independently and must be addressed simultaneously via transformer design \cite{Yin2021ImpactConverter}.
  \item Supply chain – electrical steel and copper have experienced severe market volatility and shortages.
\end{enumerate}

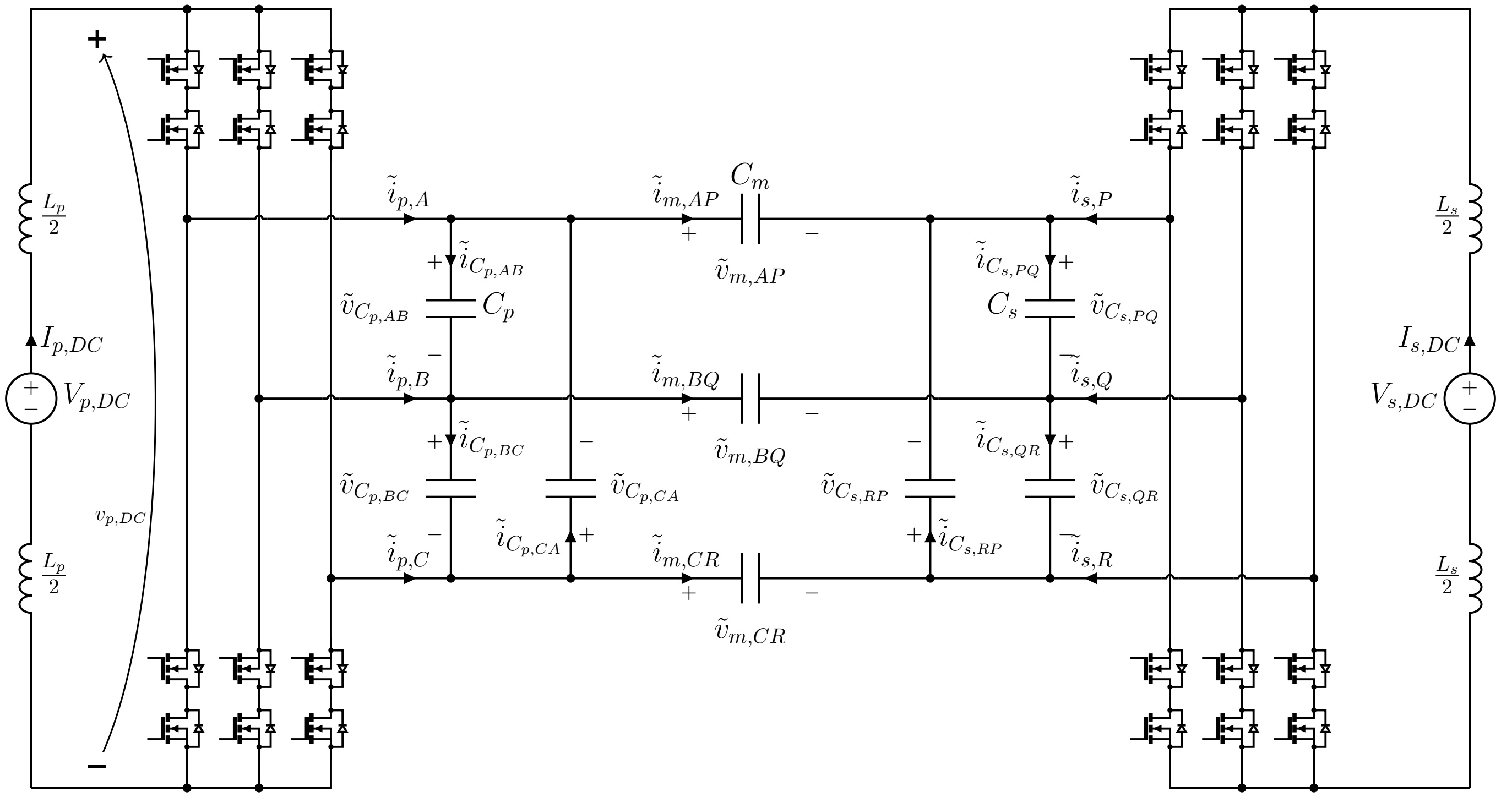
\begin{figure*}
\trimbox{0.35cm 0cm 0.35cm 0cm}{ 
\scalebox{0.72}{
\begin{circuitikz}[american,voltage dir=old,line width=1pt]
\ctikzset{diodes/scale=0.7}
\ctikzset{bipoles/thickness=1}
\ctikzset{bipole label style/.style={font=\Large}}
\ctikzset{bipole voltage style/.style={font=\Large}}
\ctikzset{bipole current style/.style={font=\Large}}
\draw           
        (0,0)   coordinate(start)
                to [short]++(0,2) to [L,l_=$\frac{L_p}{2}$] ++ (0,3)
                to [V,v=$ $,l_=$V_{p,DC}$,i_>=$I_{p,DC}$]++(0,3)
                to [L,l_=$\frac{L_p}{2}$] ++ (0,3) to [short]++(0,2)
                to [short, -*] ++ (2.6,0) coordinate(a1)
                to [short, -*] ++ (1.2,0) coordinate(b1)
                to [short] ++ (1.2,0) coordinate(c1)
                to [short] ++ (0,-0.5)
                node[nigfete, bodydiode, anchor=S, yscale=-0.66, xscale=0.66] (PHTC) {} ++ (0,-1)
                node[nigfete, bodydiode, anchor=D, yscale=0.66, xscale=0.66] (PHBC) {} ++ (0,-1)
                to [short, -*] ++ (0,-7) coordinate (cp) 
                to [short] ++ (0,-1)
                node[nigfete, bodydiode, anchor=S, yscale=-0.66, xscale=0.66] (PLTC) {} ++ (0,-1)
                node[nigfete, bodydiode, anchor=D, yscale=0.66, xscale=0.66] (PLBC) {} ++ (0,-1)
                to [short] ++ (0,-0.5)
                to [short, -*] ++ (-1.2, 0)
                to [short, -*] ++ (-1.2, 0)
                to [short] ++ (-2.6, 0) 
        (b1)    to [short] ++ (0,-0.5)
                node[nigfete, bodydiode, anchor=S, yscale=-0.66, xscale=0.66] (PHTB) {} ++ (0,-1)
                node[nigfete, bodydiode, anchor=D, yscale=0.66, xscale=0.66] (PHBB) {} ++ (0,-1)
                to [short, -*] ++ (0,-4) coordinate (bp) 
                to [short] ++ (0,-4)
                node[nigfete, bodydiode, anchor=S, yscale=-0.66, xscale=0.66] (PLTB) {} ++ (0,-1)
                node[nigfete, bodydiode, anchor=D, yscale=0.66, xscale=0.66] (PLBB) {} ++ (0,-1)
                to [short] ++ (0,-0.5)
        (a1)    to [short] ++ (0,-0.5)
                node[nigfete, bodydiode, anchor=S, yscale=-0.66, xscale=0.66] (PHTA) {} ++ (0,-1)
                node[nigfete, bodydiode, anchor=D, yscale=0.66, xscale=0.66] (PHBA) {} ++ (0,-1)
                to [short, -*] ++ (0,-1) coordinate (ap) 
                to [short] ++ (0,-7)
                node[nigfete, bodydiode, anchor=S, yscale=-0.66, xscale=0.66] (PLTA) {} ++ (0,-1)
                node[nigfete, bodydiode, anchor=D, yscale=0.66, xscale=0.66] (PLBA) {} ++ (0,-1)
                to [short] ++ (0,-0.5)

        (ap)    to [short]++(0.6,0) to [crossing]++(1.2,0) to [crossing]++(1.2,0)
                to [short,i=$\tilde{i}_{p,A}$]++(1.4,0) coordinate(f1) to [short]++(3,0)
                to [C,v=$\tilde{v}_{m,AP}$,i>^=$\tilde{i}_{m,AP}$,l=$C_m$]++(4,0)
                to [short]++(3,0) coordinate(f2) to [short,i<=$\tilde{i}_{s,P}$]++(1.4,0)
                to [short]++(0.6,0)
                
        (bp)    to [short]++(0.6,0) to [crossing]++(1.2,0)
                to [short,i=$\tilde{i}_{p,B}$]++(1.4,0) to [short]++(1,0) to [crossing]++(2,0)
                to [C,v=$\tilde{v}_{m,BQ}$,i>^=$\tilde{i}_{m,BQ}$]++(4,0)
                to [crossing]++(2,0) to [short]++(1,0) to [short,i<=$\tilde{i}_{s,Q}$]++(1.4,0)
                to [crossing]++(1.2,0) to [short]++(0.6,0)
                
        (cp)    to [short]++(0.6,0)
                to [short,i=$\tilde{i}_{p,C}$]++(1.4,0) to [short]++(3,0)
                to [C,v=$\tilde{v}_{m,CR}$,i>^=$\tilde{i}_{m,CR}$]++(4,0) 
                to [short]++(3,0) to [short,i<=$\tilde{i}_{s,R}$]++(1.4,0)
                to [crossing]++(1.2,0) to [crossing]++(1.2,0) to [short]++(0.6,0)

        (f1)    to [C,v=$\tilde{v}_{C_{p,AB}}$,i>^=$\tilde{i}_{C_{p,AB}}$,l=$C_p$,*-*]++(0,-3)
                to [C,v=$\tilde{v}_{C_{p,BC}}$,i>^=$\tilde{i}_{C_{p,BC}}$,-*]++(0,-3)
                ++(2,0) to [C,v=$\tilde{v}_{C_{p,CA}}$,i>^=$\tilde{i}_{C_{p,CA}}$,*-]++(0,3) to [short,-*]++(0,3)

        (f2)    to [C,v^=$\tilde{v}_{C_{s,PQ}}$,i>_=$\tilde{i}_{C_{s,PQ}}$,l_=$C_s$,*-*]++(0,-3)
                to [C,v^=$\tilde{v}_{C_{s,QR}}$,i>_=$\tilde{i}_{C_{s,QR}}$,-*]++(0,-3)
                ++(-2,0) to [C,v^=$\tilde{v}_{C_{s,RP}}$,i>_=$\tilde{i}_{C_{s,RP}}$,*-]++(0,3) to [short,-*]++(0,3)

        (24,0)  coordinate (start2)
                to [short]++(0,2) to [L,l=$\frac{L_s}{2}$,mirror]++(0,3)
                to [V,v>=$ $,l=$V_{s,DC}$,i^=$I_{s,DC}$]++(0,3)
                to [L,l=$\frac{L_s}{2}$,mirror]++(0,3) to [short]++(0,2)
                to [short, -*] ++ (-2.6,0) coordinate(c2)
                to [short, -*] ++ (-1.2,0) coordinate(b2)
                to [short] ++ (-1.2,0) coordinate(a2)
                to [short] ++ (0,-0.5)
                node[nigfete, bodydiode, anchor=S, yscale=-0.66, xscale=0.66] (SHTA) {} ++ (0,-1)
                node[nigfete, bodydiode, anchor=D, yscale=0.66, xscale=0.66] (SHBA) {} ++ (0,-1)
                to [short, -*] ++ (0,-1) coordinate (ap) 
                to [short] ++ (0,-7)
                node[nigfete, bodydiode, anchor=S, yscale=-0.66, xscale=0.66] (SLTA) {} ++ (0,-1)
                node[nigfete, bodydiode, anchor=D, yscale=0.66, xscale=0.66] (SLBA) {} ++ (0,-1)
                to [short] ++ (0,-0.5)
                to [short, -*] ++ (1.2, 0)
                to [short, -*] ++ (1.2, 0)
                to [short] ++ (2.6, 0)                
        (b2)    to [short] ++ (0,-0.5)
                node[nigfete, bodydiode, anchor=S, yscale=-0.66, xscale=0.66] (SHTB) {} ++ (0,-1)
                node[nigfete, bodydiode, anchor=D, yscale=0.66, xscale=0.66] (SHBB) {} ++ (0,-1)
                to [short, -*] ++ (0,-4) coordinate (bp) 
                to [short] ++ (0,-4)
                node[nigfete, bodydiode, anchor=S, yscale=-0.66, xscale=0.66] (SLTB) {} ++ (0,-1)
                node[nigfete, bodydiode, anchor=D, yscale=0.66, xscale=0.66] (SLBB) {} ++ (0,-1)
                to [short] ++ (0,-0.5)
        (c2)    to [short] ++ (0,-0.5)
                node[nigfete, bodydiode, anchor=S, yscale=-0.66, xscale=0.66] (SHTC) {} ++ (0,-1)
                node[nigfete, bodydiode, anchor=D, yscale=0.66, xscale=0.66] (SHBC) {} ++ (0,-1)
                to [short, -*] ++ (0,-7) coordinate (cp) 
                to [short] ++ (0,-1)
                node[nigfete, bodydiode, anchor=S, yscale=-0.66, xscale=0.66] (SLTC) {} ++ (0,-1)
                node[nigfete, bodydiode, anchor=D, yscale=0.66, xscale=0.66] (SLBC) {} ++ (0,-1)
                to [short] ++ (0,-0.5)
        ;


        


        

        \draw (1.1,0.3) node {\LARGE \textbf{--}};
        \draw (1.1,12.5) node {\LARGE \textbf{+}};
        \draw (1.5,4.5) node {$v_{p,DC}$};
        \draw [->, thick, rotate=0] (1.2,0.6) arc [start angle=-59, end angle=59, x radius=1.8, y radius=6.8];
\end{circuitikz}
}
}
\captionsetup{font={color=black, footnotesize}, belowskip=-4pt}
\caption{Proposed three-phase Capacitive Bridge Transformer topology, with primary and secondary CSI filter caps in $\Delta$-connection. This circuit diagram shows anti-series MOSFETS, but the CBT hardware should be realized with monolithic GaN four-quadrant switches for minimum conduction losses and reverse voltage blocking.}
\label{fig:big-cbt-circuit}
\end{figure*}

\color{black}To overcome these challenges, the conference paper version of this article \cite{Nowakowski2025CBT} proposed a new DAB architecture that replaces the conventional transformer with a capacitive bridge to provide galvanic isolation. \color{black} Capacitors are more energy dense than magnetics \cite{Brooks2024OnDesigns}, scale well with frequency, have lower series resistance, and suffer less market volatility. However, the capacitive heart of the architecture requires a pivot away from VSIs since they cannot drive capacitive loads directly. \color{black}Hence\color{black}, two current source inverters (CSIs) are joined at their AC terminals by a capacitive bridge, eliminating all AC magnetic components. This new DAB converter is termed a Capacitive Bridge Transformer (CBT) and is shown in Fig. \ref{fig:big-cbt-circuit}. \color{black} The following sections: 1. Examine the prior art, 2. Derive the CBT DC-DC power flow equations, effective turns ratio, and internal apparent power, 3. Verify these analytical models through measurements on an experimental prototype, and 4. discuss ongoing work.   \color{black}

\section{CBT Duality and Prior Art}
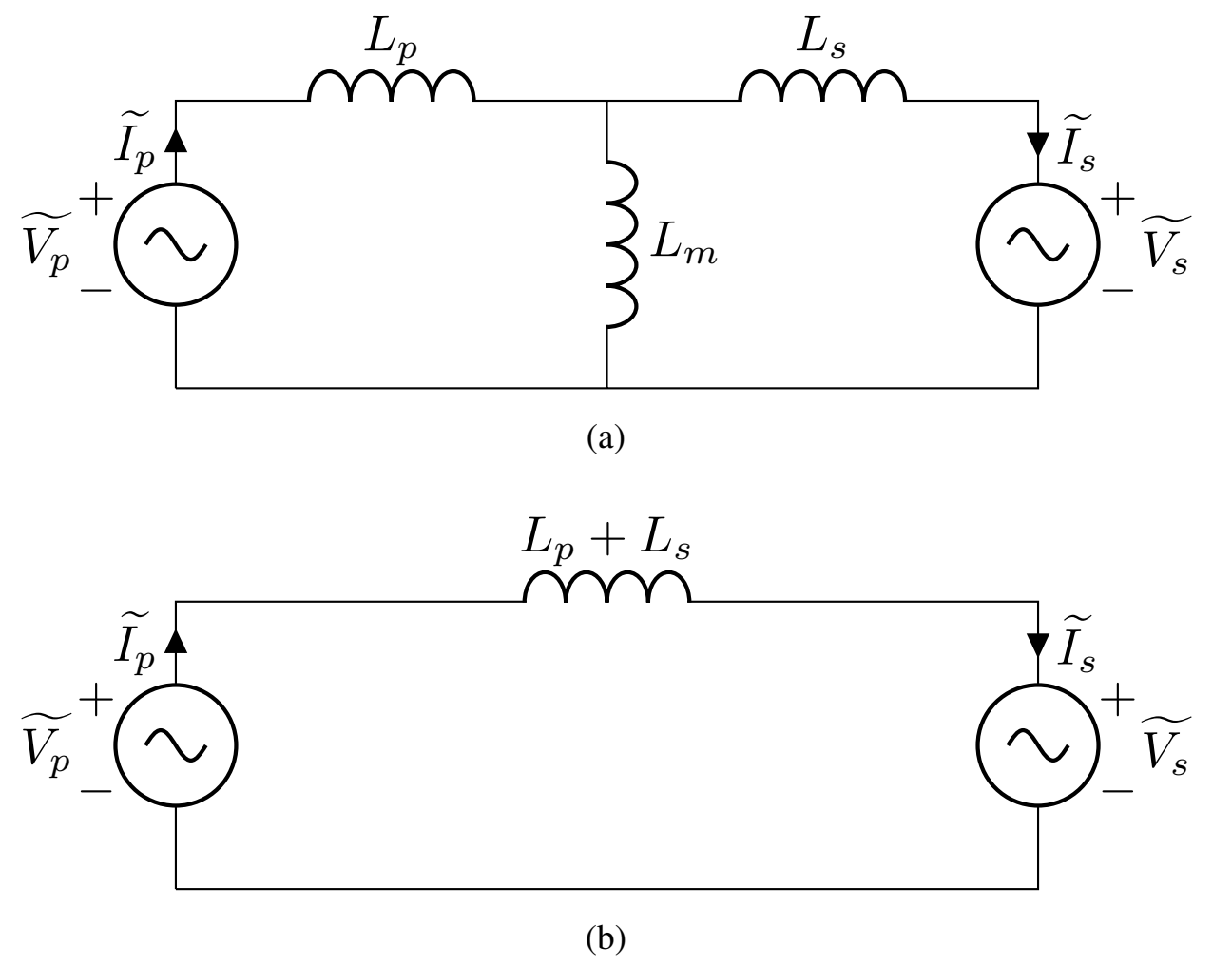
\begin{figure}[htp]

\begin{subfigure}{0.5\textwidth}
\centering
\trimbox{0.35cm 0cm 0.35cm 0cm}{
\scalebox{1.0}{
\begin{circuitikz}[american,voltage dir=old]
\draw           
        (0,0)   coordinate(start)
                to [sV,v<=$\widetilde{V_{p}}$,i>=$\widetilde{I_{p}}$]++(0,2)
                to [L,l=$L_{p}$]++(3,0)

                to [L,l=$L_{s}$]++(3,0)
                to [sV,v=$\widetilde{V_{s}}$,i>^=$\widetilde{I_{s}}$]++(0,-2)
                to (start)

        (3,2)   to [L,l=$L_m$]++(0,-2)
        ;
\end{circuitikz}
}
}
\caption{}
\end{subfigure}

\bigskip

\begin{subfigure}{0.5\textwidth}
\centering
\trimbox{0.35cm 0cm 0.35cm 0cm}{
\scalebox{1.0}{
\begin{circuitikz}[american,voltage dir=old]
\draw           
        (0,0)   coordinate(start)
                to [sV,v<=$\widetilde{V_{p}}$,i>=$\widetilde{I_{p}}$]++(0,2)
                to [short]++(1.5,0) to [L,l=$L_p+L_s$]++(3,0) to [short]++(1.5,0) 
                to [sV,v=$\widetilde{V_{s}}$,i>^=$\widetilde{I_{s}}$]++(0,-2)
                to (start)
        ;
\end{circuitikz}
}
}
\caption{}
\end{subfigure}

\caption{Single-phase models for magnetic transformer. (a) The classic magnetic transformer "T" circuit. (b) When $L_m >> L_p, L_s$, the single-phase magnetic transformer can be represented as a series circuit without $L_m$.}
\label{fig:magnetic-tfmr-circuit}
\end{figure}

The CBT concept is best introduced through duality with its magnetic DAB counterpart on a per-phase basis, as shown in Fig. \ref{fig:magnetic-tfmr-circuit} and Fig. \ref{fig:capacitive-tfmr-circuit}.  Fig. \ref{fig:magnetic-tfmr-circuit}(a) represents the classic “T” circuit of the transformer, which simplifies to Fig. \ref{fig:magnetic-tfmr-circuit}(b) when magnetizing inductance is far bigger than the leakages, i.e. $L_m >> L_p, L_s$.  The VSIs are represented by ideal AC voltage sources. The CBT is shown in Fig. \ref{fig:capacitive-tfmr-circuit}(a), where capacitors replace the transformer in a “$\pi$” circuit. The CSIs are represented by ideal AC current sources. The mutual capacitance $C_m$ provides isolation and when $C_m >> C_p, C_s$, simplifies the circuit to Fig. \ref{fig:capacitive-tfmr-circuit}(b) from a power flow analysis perspective. However, unlike the transformer version where $L_p$, $L_s$, $L_m$ are lumped elements representing a single physical component, i.e. the transformer, the CBT capacitors $C_p$, $C_s$, $C_m$ are truly discrete elements decoupled from each other and their values may be selected independently at the designer's discretion. 

The non-isolated circuit in Fig. \ref{fig:capacitive-tfmr-circuit}(b) was introduced in \cite{Peterson1978US4079305ALoads}, where two superconducting magnetic energy storage (SMES) coils served as the DC link inductors of two thyristor CSIs. Energy was exchanged between the coils using firing angle control with capacitive commutation on the AC terminal link. Although capacitors were the only passive AC components, there was no galvanic isolation. Interestingly, this current source DAB predates the well-known voltage source incarnation \cite{Doncker1991AApplications} by at least 15 years. After its patenting, the current source DAB circuit was examined for niche high-power superconducting physics experiments during the 1980s \cite{Kustom1980ThyristorCoils,Fuja1980Three-PhaseCoils,Ehsani1985AMagnets} and then for general purpose DC-DC applications in the early 1990s \cite{Ehsani1993InverseApplications,Bilgic2003AnalysisStorage}, but it never gained widespread use.  Lack of adoption likely stemmed from its non-isolated nature and the use of thyristor CSIs switching at line frequency, resulting in very large passive components and low power density.  Replacing the thyristors with anti-series connected IGBTs or MOSFETs could have allowed higher switching frequencies but with significant efficiency and cost penalties due to twice the device count.  That penalty, however, is no longer a barrier as new monolithic bidirectional blocking devices become commercially available \cite{Baliga2023TheConverters}, leaving only the matter of isolation.  

Modern efforts in wireless capacitive power transfer (CPT) and multilevel converters have prompted detailed mutual capacitive coupling models \cite{Zhang2016AApplication, LudoisDai2015Survey}.  Kilowatt scale CPT was first demonstrated in \cite{Dai2015SingleTransfer} with buck, boost, buck-boost and SEPIC circuits and later expanded to higher powers and isolation voltage in \cite{Kamalapur2025DimensionalConverters}. High voltage DC-DC converters formed by the series connection of capacitively isolated rectifiers was shown in \cite{Channegowda2020HighCharacteristics, Neuner2023AnalysisTransformer}, but still utilize AC magnetics with VSI building blocks. In all these scenarios isolation is achieved by adding or exploiting series capacitance in the AC portions of the circuit.  The marriage of these isolation concepts with the CSI DAB is how capacitive isolation is shown in this paper as Fig. \ref{fig:capacitive-tfmr-circuit}(a). Although prior art derived the basic CSI DAB power flow equation \cite{Kustom1980ThyristorCoils,Fuja1980Three-PhaseCoils}, the expansion of the AC capacitive link from a single three-phase capacitor set into an isolated three-phase bridge of capacitors requires renewed attention.  In the following section, the new power flow equations for the CBT are derived in detail. 

\section{CBT Power Flow Analysis} \label{sec:model}

\begin{figure}[htp]

\begin{subfigure}{0.5\textwidth}
\centering
\trimbox{0.35cm 0cm 0.35cm 0cm}{
\scalebox{0.95}{
\begin{circuitikz}[american,voltage dir=old]
\draw           
        (0,0)   coordinate(start1)
                to [isource,l=$\widetilde{i_{p}}$]++(0,3)
                to [short]++(2,0) node[label={[font=\footnotesize]above:A}] {}
                to [C,i>^=$\tilde{i}_{C_p}$,v=$v_{C_p}$,l^=$C_p$]++(0,-3)
                node[label={[font=\footnotesize]below:B}] {} to [short] (start1)
                
        (7,0)   coordinate(start2)
                to [isource,l_=$\widetilde{i_{s}}$]++(0,3)
                to [short]++(-2,0) node[label={[font=\footnotesize]above:C}] {}
                to [C,i>^=$\tilde{i}_{C_s}$,l_=$C_s$]++(0,-3)
                node[label={[font=\footnotesize]below:D}] {} to [short] (start2)
                
        (2,3)   to [C,i>^=$\tilde{i}_{C_m}$,l_=$C_m$,*-*]++(3,0)
        (2,0)   to [C,i>_=$\tilde{i}_{C_m'}$,l=$C_m'$,*-*]++(3,0)
        ;
\end{circuitikz}
}
}
\caption{}
\end{subfigure}

\bigskip

\begin{subfigure}{0.5\textwidth}
\centering
\trimbox{0.35cm 0cm 0.35cm 0cm}{
\scalebox{1.1}{
\begin{circuitikz}[american,voltage dir=old]
\draw           
        (0,0)   coordinate(start)
                to [isource,l=$\widetilde{i_{p}}$]++(0,2)
                to [short]++(6,0)
                to [isource,l=$\widetilde{i_{s}}$,invert]++(0,-2)
                to (start)
        
        (3,2) to [C,l=$C_p+C_s$]++(0,-2)
        ;
\end{circuitikz}
}
}
\caption{}
\end{subfigure}

\caption{Single-phase models for capacitive transformer. (a) The capacitive transformer "$\pi$" circuit. (b) When $C_m >> C_p, C_s$, the single-phase capacitive transformer can be represented as a parallel circuit without $C_m$.}
\label{fig:capacitive-tfmr-circuit}
\end{figure}
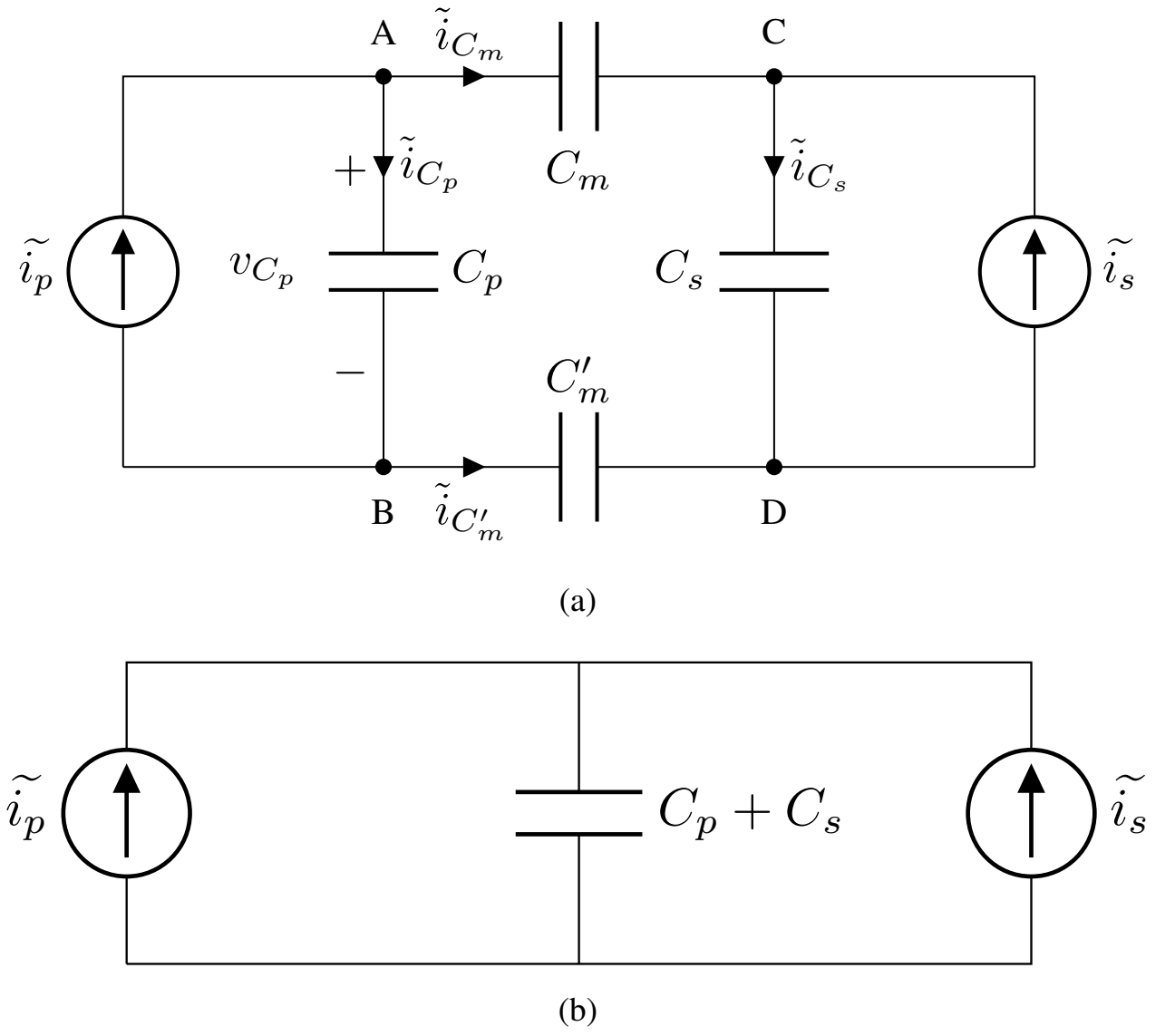

To systematically understand the circuit operation of the capacitive bridge transformer topology featuring an isolated bridge of capacitors, \color{black}the power flow is first derived for a single-phase CBT with ideal sinusoidal current sources, as shown by the topology in Fig. \ref{fig:capacitive-tfmr-circuit}(a). Using a Thévenin equivalent, the power sourced by the primary side is given in \eqref{eq:P_ip_1ph_sine}. For this equation, $\theta$ is the phase shift between the primary and secondary-side sources, $I_p$ and $I_s$ are RMS quantities, and $\omega$ is the frequency in radians per second. Equation \eqref{eq:P_ip_1ph_sine} shows that the power flow for the CBT topology can be defined in terms of the two source magnitudes, the phase shift between them, and the joining impedance, exactly like a VSI-DAB topology or electric machine equivalent circuit.


\color{black}
\begin{align}
P (\theta) &= \frac{I_p^{rms}I_s^{rms}\sin (\theta)}{\omega C_pC_s\left(\frac{1}{C_p}+\frac{1}{C_m}+\frac{1}{C_m'}+\frac{1}{C_s}\right)}\label{eq:P_ip_1ph_sine}
\end{align}
\color{black}

\color{black} However, a true implementation of a CBT circuit utilizing CSIs will have chopped rectangular currents, significantly complicating the calculations but will retain the same physical operation.  To simplify the analysis for the three-phase CBT with delta-connected filter capacitors matching the topology presented in Fig. \ref{fig:big-cbt-circuit}, the real power flow, internal apparent power, and the voltage/current transfer ratio expressions are derived in a step-by-step manner exploiting the concept of symmetry. The derivations lay emphasis on a 1/6 cycle which can be extended to the entire periodic cycle. $\Phi$ is defined as the normalized phase shift angle, where $2\pi$ represents unity. \color{black} A closer analysis reveals that varying the phase-shift between the two CSI bridges results in the CBT topology being reduced to one of the two equivalent circuits based on the switching states. Further, based on the range of phase-shift magnitude, the operational analysis can be conducted in three cases. Section \ref{sec:case1} discusses the case when $0<\Phi<1/6$, Section \ref{sec:case2} discusses the case when $1/6<\Phi<1/3$ and Section \ref{sec:case3} discusses the case when $1/3<\Phi<1/2$. Together, the three cases cover the complete power flow analysis in one direction and the analysis can simply be extended for regenerative power flow operation.


\subsection{Case 1: $0<\Phi<1/6$} \label{sec:case1}

\begin{figure}[htbp]
\centerline{\includegraphics[width=0.9\columnwidth]{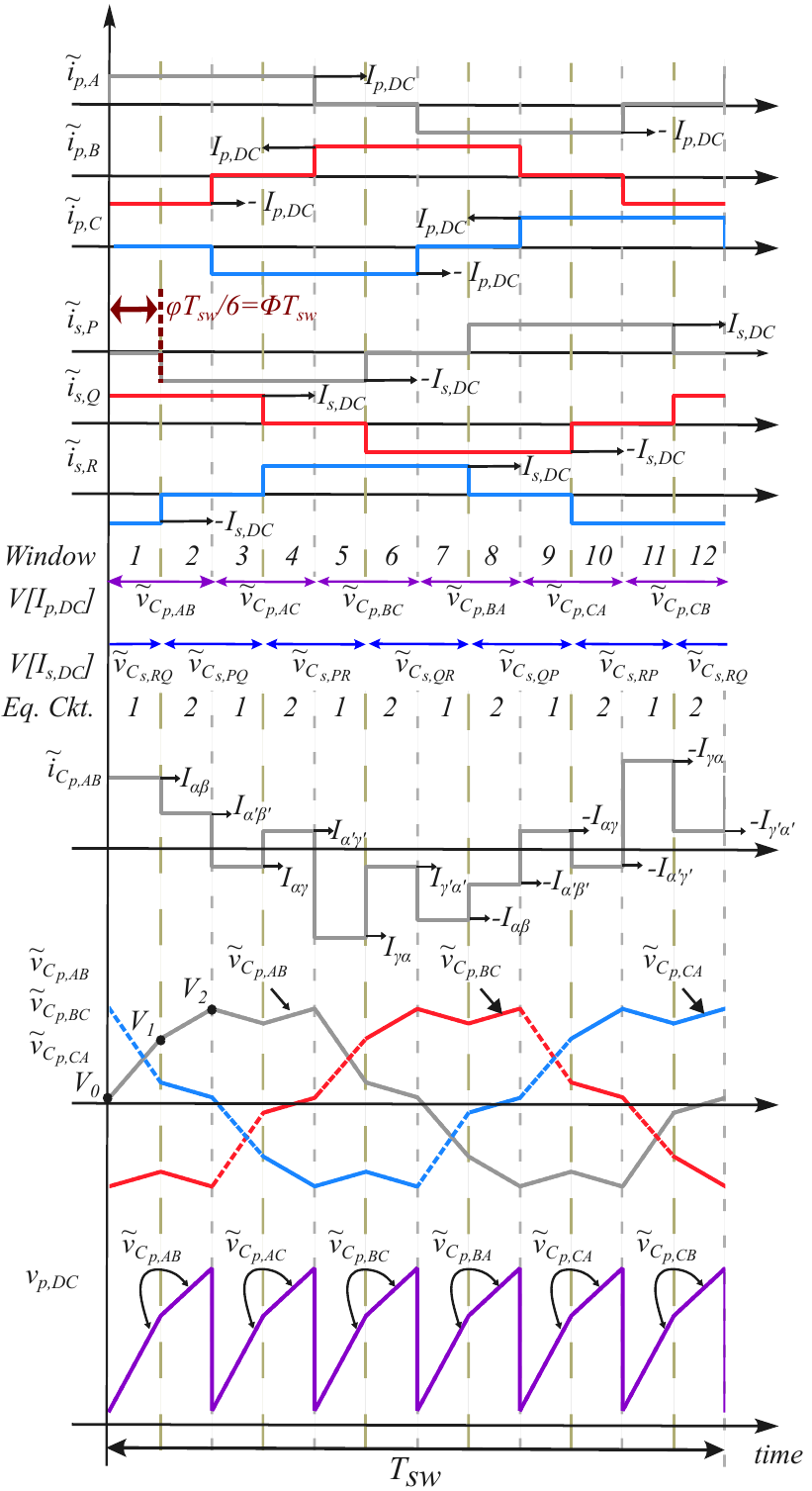}}
\caption{Various analytical waveforms of the capacitive bridge transformer topology depicting the operating method when $0<\Phi<1/6$.}
\label{fig:waveforms_3Phase1}
\end{figure}

Fig.~\ref{fig:waveforms_3Phase1} qualitatively illustrates a few underlying analytical waveforms that depict the main operating method of a CBT topology. The primary and secondary side CSI bridges, which are phase-shifted by $\Phi$, operate under six-step modulation. The current waveforms at the outer ports of each of the CSI bridges are presented in the top few windows of the figure where, $I_{p,x}$ and $I_{s,y}$ represent the primary and secondary side throw currents and $x=A,B,C$ and $y=P,Q,R$ respectively. While the small-dashed vertical lines indicate the $\pi/3$ sector division, the big-dashed vertical lines indicate the phase-shift, $\Phi$, between the bridges. Using symmetry, the complete switching cycle can be divided into twelve windows as shown in the figure. Based on the window of operation, the voltages across the delta capacitors $C_{p,xx'}$ and $C_{s,yy'}$ manifest across the current-stiff ports $I_{p,DC}$ and $I_{s,DC}$ respectively, also noted in the figure as $V[I_{p,DC}]$ and $V[I_{s,DC}]$. The remaining section derives the power flow equation in a step-by-step manner by considering half-cycle operation due to symmetry. 

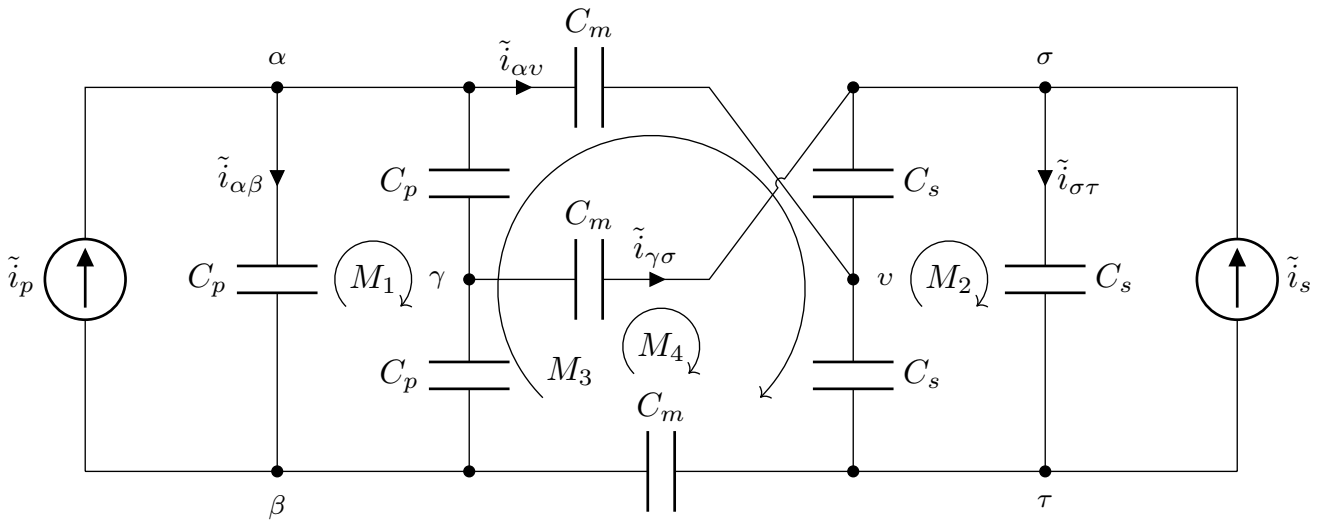
\begin{figure}[htbp]
\centering
\trimbox{0.36cm 0.12cm 0.36cm 0.12cm}{ 
\scalebox{0.635}{
\begin{circuitikz}[american,voltage dir=old]
\draw       
    (0,0)   coordinate(start) 
            to [short]++(0,1) to [I,l=$\tilde{i}_p$]++(0,2) to [short]++(0,1)
            to [short]++(2,0) node[label={[font=\footnotesize]above:$\alpha$}] {} coordinate(c1) to [short]++(2,0)
            coordinate(b11) to [C,l_=$C_p$]++(0,-2)  coordinate(b12) node[label={[font=\footnotesize]left:$\gamma$}] {} to [C,l_=$C_p$]++(0,-2)  coordinate(b13)
            to [short]++(-2,0) node[label={[font=\footnotesize]below:$\beta$}] {} to [short]++(-2,0)
    (c1)    to [C,l_=$C_p$,i>_=$\tilde{i}_{\alpha \beta}$,*-*]++(0,-4)

    (12,0)  coordinate(start2) 
            to [short]++(0,1) to [I,l_=$\tilde{i}_s$]++(0,2) to [short]++(0,1)
            to [short]++(-2,0) node[label={[font=\footnotesize]above:$\sigma$}] {} coordinate(c2) to [short]++(-2,0)
            coordinate(b21) to [C,l=$C_s$]++(0,-2)  coordinate(b22) node[label={[font=\footnotesize]right:$\upsilon$}] {} to [C,l=$C_s$]++(0,-2)  coordinate(b23)
            to [short]++(2,0) node[label={[font=\footnotesize]below:$\tau$}] {} to [short]++(2,0)
    (c2)    to [C,l=$C_s$,i>^=$\tilde{i}_{\sigma \tau}$,*-*]++(0,-4)

    (b11)   to [C,l=$C_m$,i>^=$\tilde{i}_{\alpha \upsilon}$,*-]++(2.5,0) to [short,-*]++(1.5,-2)
    (b12)   to [C,l=$C_m$,i^>=$\tilde{i}_{\gamma \sigma}$,*-]++(2.5,0) to [crossing,-*]++(1.5,2)
    (b13)   to [C,l=$C_m$,*-*]++(4,0)
        ;
    \draw[->,shift={(3,2)}] (225:.4cm) arc (225:-45:.4cm) node at(0,0){$M_1$};
    \draw[->,shift={(9,2)}] (225:.4cm) arc (225:-45:.4cm) node at(0,0){$M_2$};
    \draw[->,shift={(5.05,1.05)}] (225:.4cm) arc (225:-45:1.6cm) node at(0,0){$M_3$};
    \draw[->,shift={(6,1.3)}] (225:.4cm) arc (225:-45:.4cm) node at(0,0){$M_4$};
\end{circuitikz}
}
}
\caption{Circuit diagram of Equivalent Circuit 1 valid for Cases 1-3.}
\label{fig:eq_ckt1}
\end{figure}

Let us consider the case when $0<\Phi<1/
$, also illustrated qualitatively in Fig.~\ref{fig:waveforms_3Phase1}. During Window 1, 3, \& 5 of operation, the ideal equivalent three-phase circuit can be represented by a simplified circuit also shown in Fig.~\ref{fig:eq_ckt1}. Here, the three primary side phases $A, B, C$ can be mapped into $\alpha, \beta, \gamma$ based on the window under consideration. Similarly, the three secondary side three phases $P, Q, R$ can be mapped into $\sigma, \tau, \upsilon$ based on the window under consideration. For example, for Window 1 of operation, $\alpha=A$, $\beta=B$, $\gamma=C$ and $\sigma=R$, $\tau=Q$, $\upsilon=P$. 

Applying KVL across Mesh $M_1-M_4$, (\ref{eq:A_m1})-(\ref{eq:A_m4}) can be derived,
\begin{align}
\tilde{i}_{p} &= + \tilde{i}_{\alpha\upsilon} + \frac{\tilde{i}_{\gamma\sigma}}{2} + \frac{3\tilde{i}_{\alpha\beta}}{2} \label{eq:A_m1} \\
\tilde{i}_{s} &= - \frac{\tilde{i}_{\alpha\upsilon}}{2} - \tilde{i}_{\gamma\sigma} + \frac{3\tilde{i}_{\sigma\tau}}{2} \label{eq:A_m2} 
\end{align}
\begin{multline}
\frac{2\tilde{i}_{p}}{C_p} - \frac{\tilde{i}_{s}}{C_s} = [\frac{2}{C_p}+\frac{2}{C_m}+\frac{1}{C_s}]\tilde{i}_{\alpha\upsilon} \\ + [\frac{1}{C_p}+\frac{1}{C_m}+\frac{1}{C_s}]\tilde{i}_{\gamma\sigma} + \frac{2\tilde{i}_{\alpha\beta}}{C_p} - \frac{\tilde{i}_{\sigma\tau}}{C_s}\label{eq:A_m3}
\end{multline}
\begin{multline}
\frac{\tilde{i}_{p}}{C_p} - \frac{2\tilde{i}_{s}}{C_s} = [\frac{1}{C_p}+\frac{1}{C_m}+\frac{1}{C_s}]\tilde{i}_{\alpha\upsilon} \\ + [\frac{1}{C_p}+\frac{2}{C_m}+\frac{2}{C_s}]\tilde{i}_{\gamma\sigma} + \frac{\tilde{i}_{\alpha\beta}}{C_p} - \frac{2\tilde{i}_{\sigma\tau}}{C_s}\label{eq:A_m4}
\end{multline}
\noindent where, $\tilde{i}_{\alpha\beta}, \tilde{i}_{\sigma\tau}, \tilde{i}_{\alpha\upsilon}, \tilde{i}_{\gamma\sigma}$ are the currents in the various delta and mutual capacitors connecting the $\alpha, \beta, \gamma$ and $\sigma, \tau, \upsilon$ nodes. Note that the currents in the remaining capacitors can simply be derived using KCL at the appropriate nodes. For example, $\tilde{i}_{\alpha\gamma}=(\tilde{i}_{p}-\tilde{i}_{\alpha\upsilon}-\tilde{i}_{\alpha\beta})$ and so on. 

Rearranging (\ref{eq:A_m1})-(\ref{eq:A_m4}), the capacitor currents can be derived in terms of the primary and secondary side currents and the capacitor values as noted in (\ref{eq:A_i1})-(\ref{eq:A_i4}).
\begin{align}
\tilde{i}_{\alpha\upsilon} &= \frac{1/C_p}{1/C_{eq}}\tilde{i}_{p} \label{eq:A_i1} \\
\tilde{i}_{\gamma\sigma} &= -\frac{1/C_s}{1/C_{eq}}\tilde{i}_{s} \label{eq:A_i2} \\
\tilde{i}_{\alpha\beta} &= \frac{2(3/C_m+1/C_s)\tilde{i}_{p}+\tilde{i}_{s}/C_s}{3/C_{eq}} \label{eq:A_i3} \\
\tilde{i}_{\sigma\tau} &=  \frac{2(1/C_p+3/C_m)\tilde{i}_{s}+\tilde{i}_{p}/C_p}{3/C_{eq}} \label{eq:A_i4}
\end{align}
\noindent where, $\frac{1}{C_{eq}}=\frac{1}{C_p}+\frac{3}{C_m}+\frac{1}{C_s}$.

\begin{figure}[htbp]
\centering
\trimbox{0.36cm 0.12cm 0.36cm 0.12cm}{ 
\scalebox{0.635}{
\begin{circuitikz}[american,voltage dir=old]
\draw       
    (0,0)   coordinate(start) 
            to [short]++(0,1) to [I,l=$\tilde{i}_p$]++(0,2) to [short]++(0,1)
            to [short]++(2,0) node[label={[font=\footnotesize]above:$\alpha '$}] {} coordinate(c1) to [short]++(2,0)
            coordinate(b11) to [C,l_=$C_p$,*-*]++(0,-2)  coordinate(b12) node[label={[font=\footnotesize]left:$\gamma '$}] {} to [C,l_=$C_p$,-*]++(0,-2)  coordinate(b13)
            to [short]++(-2,0) node[label={[font=\footnotesize]below:$\beta '$}] {} to [short]++(-2,0)
    (c1)    to [C,l_=$C_p$,i>_=$\tilde{i}_{\alpha ' \beta '}$,*-*]++(0,-4)

    (12,0)  coordinate(start2) 
            to [short]++(0,1) to [I,l_=$\tilde{i}_s$]++(0,2) to [short]++(0,1)
            to [short]++(-2,0) node[label={[font=\footnotesize]above:$\sigma '$}] {} coordinate(c2) to [short]++(-2,0)
            coordinate(b21) to [C,l=$C_s$,*-*]++(0,-2)  coordinate(b22) node[label={[font=\footnotesize]right:$\upsilon '$}] {} to [C,l=$C_s$,-*]++(0,-2)  coordinate(b23)
            to [short]++(2,0) node[label={[font=\footnotesize]below:$\tau '$}] {} to [short]++(2,0)
    (c2)    to [C,l=$C_s$,*-*]++(0,-4)

    (b11)   to [C,l=$C_m$]++(4,0);
    \draw[gray,opacity=0.7]
    (b12)   to [C,l=$C_m$,color=gray,opacity=0.7]++(4,0);
    \draw
    (b13)   to [C,l=$C_m$]++(4,0)
        ;
    \draw [->, thick, rotate=0] (4.7,0.05) arc [start angle=250, end angle=-70, x radius=3.8cm, y radius=2.5cm] node at(4.9,0.3){$M_5$};
\end{circuitikz}
}
}
\caption{Circuit diagram of Equivalent Circuit 2 valid for Cases 1, 3.}
\label{fig:eq_ckt2}
\end{figure}
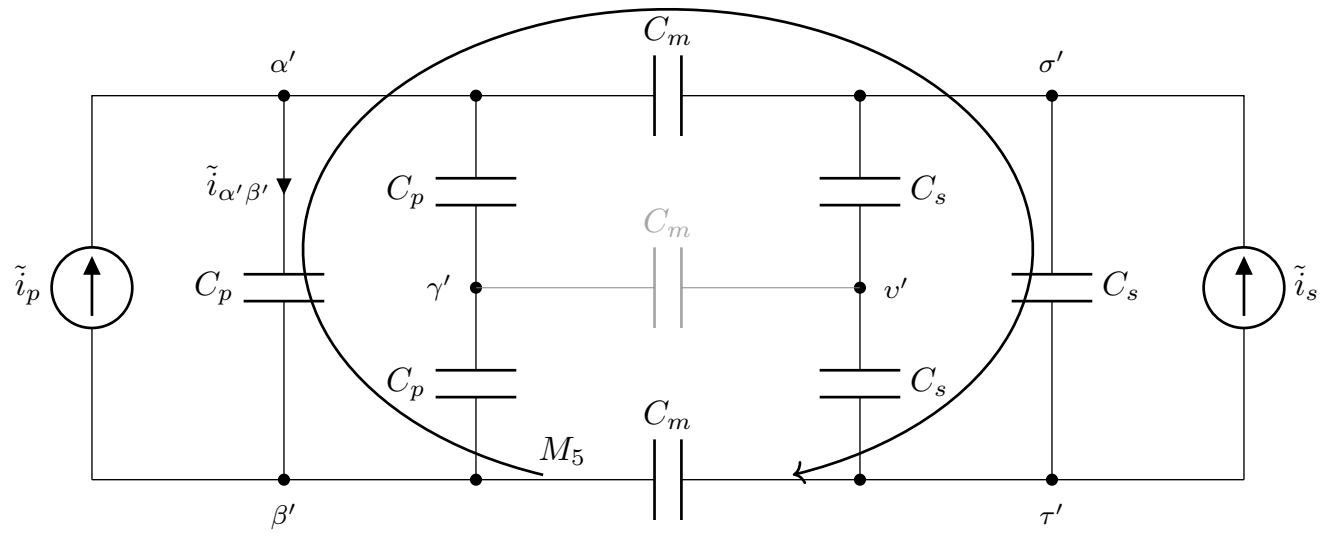

During Window 2, 4, \& 6 of operation, the ideal equivalent three-phase circuit can be represented as shown in Fig.~\ref{fig:eq_ckt2}. Here, the three primary side phases $A, B, C$ and the three secondary side three phases $P, Q, R$ can be mapped into $\alpha', \beta', \gamma'$ and $\sigma', \tau', \upsilon'$ based on the window under consideration. Note that branch $\gamma'\upsilon'$ has been grayed out since the current flow will be zero due to symmetry. Further, only current $i_{\alpha'\beta'}$ has been annotated in this circuit as the rest of the capacitor currents can simply be derived as a function of $\tilde{i}_{p}, \tilde{i}_{s}, \tilde{i}_{\alpha'\beta'}$. For example, $\tilde{i}_{\alpha'\gamma'}=\frac{\tilde{i}_{\alpha'\beta'}}{2}$ and so on.

Applying KVL across Mesh $M_5$, (\ref{eq:A_i5}) can be derived, 
\begin{align}
\tilde{i}_{\alpha'\beta'} &= \frac{2(3/C_m+1/C_s)\tilde{i}_{p}+2\tilde{i}_{s}/C_s}{3/C_{eq}} \label{eq:A_i5} 
\end{align}
The derivation of these current expressions eases the power flow derivation. From Fig.~\ref{fig:waveforms_3Phase1} it can be observed that using symmetry, the average of the voltage waveform across a delta capacitor over 1/6 cycle can yield the average DC voltage across the source or the load ports. Considering Window 1 of operation from the figure, note that $P_p=I_p \overline{v}_{C_{p,AB}}|_{T_{sw}/6}$ where, 
\begin{align}
    \overline{v}_{C_{p,AB}}|_{T_{sw}/6} &= \frac{6}{T_{sw}} \left[ \int_0^{\varphi T_{sw}}\tilde{v}_{C_{p,AB}}dt+ \int_{\varphi T_{sw}}^{ T_{sw}/6}\tilde{v}_{C_{p,AB}}dt \right] \\
    &= \frac{1}{2}\left[ \varphi V_0 + V_1 + \varphi'V_2 \right] \label{eq:A_vmean1}
\end{align}
\noindent and, $V_0, V_1$ and $V_2$ are the capacitor voltages at the beginning of Windows $1, 2$ and $3$ respectively, $\varphi$ is the per-unitized duty ratio over 1/6 cycle such that $\varphi=\Phi/6$ and, 
\begin{align}
    V_1 &= V_0 + \frac{\varphi T_{sw}}{6C_p}\tilde{i}_{C_{p,AB}}[1]\label{eq:A_vmean2} \\
    V_2 &= V_1 + \frac{\varphi T_{sw}}{6C_p}\tilde{i}_{C_{p,AB}}[2]\label{eq:A_vmean3}
\end{align}
Substituting (\ref{eq:A_i1}), (\ref{eq:A_i5}), (\ref{eq:A_vmean2}), (\ref{eq:A_vmean3}) into (\ref{eq:A_vmean1}), the averaged voltage equation can be simplified to be as noted next. 
\begin{multline}
    \overline{v}_{C_{p,AB}}|_{T_{sw}/6} = V_0 + \\ \frac{T_{sw}}{36C_p} \left[ \frac{3/C_m+1/C_s}{1/C_{eq}}2\tilde{i}_p + \frac{1/C_s}{1/C_{eq}}\tilde{i}_s(1+\varphi'^2)\right] \label{eq:A_vmean4}
\end{multline}
\noindent where, $\varphi'=1-\varphi$ and the initial steady-state capacitor voltage, $V_0$, can be calculated by assuming that the voltage averaged over the full fundamental cycle, $T_{sw}$, is zero. 
\begin{align}
    \overline{v}_{C_{p,AB}}|_{T_{sw}} &= \frac{1}{T_{sw}} \left[ \int_0^{ T_{sw}}\tilde{v}_{C_{p,AB}}dt \right] = 0
\end{align}
\begin{multline}
\implies V_0 = -\frac{T_{sw}}{12C_p}[\varphi(i_{C_{p,AB}}[1]+i_{C_{p,AB}}[3]+i_{C_{p,AB}}[5])\\+\varphi'(i_{C_{p,AB}}[2]+i_{C_{p,AB}}[4]+i_{C_{p,AB}}[6])] \label{eq:A_Vinit1}
\end{multline}
\noindent where, the capacitor currents can be mapped to current expressions derived for the equivalent circuits as noted in Table \ref{tab:map_iCAB}. Substituting the current expressions from the table, the final expression for $V_0$ can be derived as noted in (\ref{eq:A_V0}).
\begin{align}
    V_0 = -\frac{T_{sw}}{18}\frac{C_{eq}}{C_p}[(3/C_m+1/C_s)\tilde{i}_p+(1+\varphi) \tilde{i}_s/C_s] \label{eq:A_V0}
\end{align}

\begin{table*}[htbp]
\caption{Mapping of the current across the delta capacitor $C_{p,AB}$ to Equivalent Circuit 1 and 2 current waveforms from Fig.~\ref{fig:eq_ckt1} and Fig.~\ref{fig:eq_ckt2} for half of the fundamental period.}
\centering
\renewcommand{\arraystretch}{1.25} 
\begin{tabular}{c|c|c|c||c|c|c||c|c|c}
\hline
\multirow{2}*{W} & \multicolumn{3}{c||}{$0<\Phi<1/6$} & \multicolumn{3}{c||}{$1/6<\Phi<1/3$} & \multicolumn{3}{c}{$1/3<\Phi<1/2$} \\  \cline{2-10}
& \textbf{Ckt.} & \textbf{Current} & \textbf{$\tilde{i}_{p}, \tilde{i}_{p} \in$}  & \textbf{Ckt.} & \textbf{Current} & \textbf{$\tilde{i}_{p}, \tilde{i}_{p} \in$} & \textbf{Ckt.} & \textbf{Current} & \textbf{$\tilde{i}_{p}, \tilde{i}_{p} \in$} \\
\hline
1 & 1 & $\tilde{i}_{\alpha\beta}$ & \multirow{6}*{$I_{p,DC}, -I_{s,DC}$} & 1 & $\tilde{i}_{\alpha\beta}$ & $I_{p,DC}, I_{s,DC}$ & 2 & $\tilde{i}_{\alpha'\beta'}$ & \multirow{6}*{$I_{p,DC}, I_{s,DC}$} \\ 
2 & 2 & $\tilde{i}_{\alpha'\beta'}$ & & 1 & $\tilde{i}_{\alpha\beta}$ & $I_{p,DC}, -I_{s,DC}$ & 1 & $\tilde{i}_{\alpha\beta}$ & \\ 
3 & 1 & $\tilde{i}_{\alpha\gamma}$ & & 1 & $\tilde{i}_{\alpha\gamma}$ & $I_{p,DC}, I_{s,DC}$ & 2 & $\tilde{i}_{\alpha'\gamma'}$ & \\  
4 & 2 & $\tilde{i}_{\alpha'\gamma'}$ & & 1 & $\tilde{i}_{\alpha\gamma}$ & $I_{p,DC}, -I_{s,DC}$ & 1 & $\tilde{i}_{\alpha\gamma}$ & \\  
5 & 1 & $\tilde{i}_{\gamma\alpha}$ & & 1 & $\tilde{i}_{\gamma\alpha}$ & $-I_{p,DC}, I_{s,DC}$ & 2 & $\tilde{i}_{\gamma'\alpha'}$ & \\ 
6 & 2 & $\tilde{i}_{\gamma'\alpha'}$ & & 1 & $\tilde{i}_{\gamma\alpha}$ & $I_{p,DC}, -I_{s,DC}$ & 1 & $\tilde{i}_{\gamma\alpha}$ & \\ 
\hline
\hline
\end{tabular}
\label{tab:map_iCAB}
\end{table*}

Using (\ref{eq:A_vmean4}), (\ref{eq:A_V0}), the power flow expression can be derived to be as noted in (\ref{eq:A_P1}) or (\ref{eq:A_P2}) in terms of the per-unitized phase-shift $\varphi$ or complete cycle phase-shift $\Phi$ respectively.
\begin{multline}
    P_p (0<\varphi<1) = \\ -\frac{T_{sw}}{36C_pC_s}\frac{I_{p,DC}I_{s,DC}}{3/C_m+1/C_p+1/C_s}\varphi(4-\varphi) \label{eq:A_P1}
\end{multline}
\begin{multline}
    P_p (0<\Phi<1/6) = \\ -\frac{T_{sw}}{3C_pC_s}\frac{I_{p,DC}I_{s,DC}}{3/C_m+1/C_p+1/C_s}\Phi(2-3\Phi) \label{eq:A_P2}
\end{multline}

Assuming an ideal converter operation, the power flow expression can be used to derive the voltage and current transfer ratio expressions as noted in (\ref{eq:A_ratio}).
\color{black}
\begin{multline}
    0<\Phi<1/6: \frac{V_{s,DC}}{V_{p,DC}} = \frac{I_{p,DC}}{I_{s,DC}} = \\ -\frac{3C_pC_sR_s}{T_{sw}}\frac{3/C_m+1/C_p+1/C_s}{\Phi(2-3\Phi)} \label{eq:A_ratio}
\end{multline}
\color{black}

\noindent where, $R_s$ is the effective secondary side load resistance and $R_s=\frac{V_{s,DC}}{I_{s,DC}}$. 

\color{black}

Lastly, the apparent and reactive power can be calculated by using (\ref{eq:S})-(\ref{eq:Q}),
\begin{align}
    S_p = v^{rms}_{p,DC}\times i^{rms}_{p,DC} = v^{rms}_{p,DC} \times I_{p,DC} \label{eq:S}
\end{align}
\begin{align}
    Q_p = \sqrt{S_p^2-P_p^2} \label{eq:Q}
\end{align}
\noindent where, $i^{rms}_{p,DC}$ is simply $I_{p,DC}$ and $v^{rms}_{p,DC}$ is the RMS voltage across the input terminals of the primary bridge, $v_{p,DC}$, also annotated in Fig. \ref{fig:big-cbt-circuit} and characteristically shown in the bottom window of Fig. \ref{fig:waveforms_3Phase1}. The RMS magnitude can be calculated as noted in (\ref{eq:rms1})-(\ref{eq:rms3}). 

\begin{equation}
    v^{rms}_{p,DC} = \sqrt{\frac{6}{T_{sw}} \left[ \int_0^{\varphi T_{sw}/6}v^2_{p,DC}(t)dt \right]} \label{eq:rms1}
\end{equation}

\begin{multline}
    = \frac{36}{T_{sw}^2} [ \int_0^{\varphi T_{sw}}(V_0+t\frac{V_1-V_0}{\varphi T_{sw}})^2dt + \\ \int_{\varphi T_{sw}}^{T_{sw}/6}(\frac{V_1-V_2}{1-\varphi}+t\frac{V_2-V_1}{(1-\varphi)T_{sw}})^2dt]^{1/2} \label{eq:rms2}
\end{multline}

\begin{equation}
    = \sqrt{\frac{\varphi V_0(V_0+V_1)+V_1^2+(1-\varphi) V_2(V_2+V_1)}{3}}
    \label{eq:rms3}
\end{equation}

Substituting (\ref{eq:A_Vinit1}), (\ref{eq:A_vmean2}), (\ref{eq:A_vmean3}) into (\ref{eq:rms3}), the final apparent power expression is noted in (\ref{eq:S2_Case1}). 
%

\begin{multline}
    S_p (1/6<\Phi<1/3) = \\ -\frac{T_{sw}I_{p,DC}I_{s,DC}}{3(C_p+C_s)} \sqrt{\frac{I_{p,DC}^2}{108I_{s,DC}^2}+\frac{I_{p,DC}}{I_{s,DC}}s_1+s_2} \label{eq:S2_Case1} 
\end{multline}

\noindent where,

\begin{align}
    s_1 &= 4\Phi^3-\Phi^2+\frac{1}{54} \\s_2 &= -6\Phi^3+3\Phi^2+\frac{1}{108} \label{eq:S3_Case1}
\end{align}

The apparent power calculation can be used for semiconductor device and capacitor sizing along with estimation of efficiencies as a function of the phase-shift and other circuit parameters. Section \ref{sec:experiments} presents details validating the analytical models for the power calculations with the laboratory-scaled experimental setup.

\color{black}

\subsection{Case 2: $1/6<\Phi<1/3$} \label{sec:case2}

\begin{figure}[htbp]
\centerline{\includegraphics[width=0.9\columnwidth]{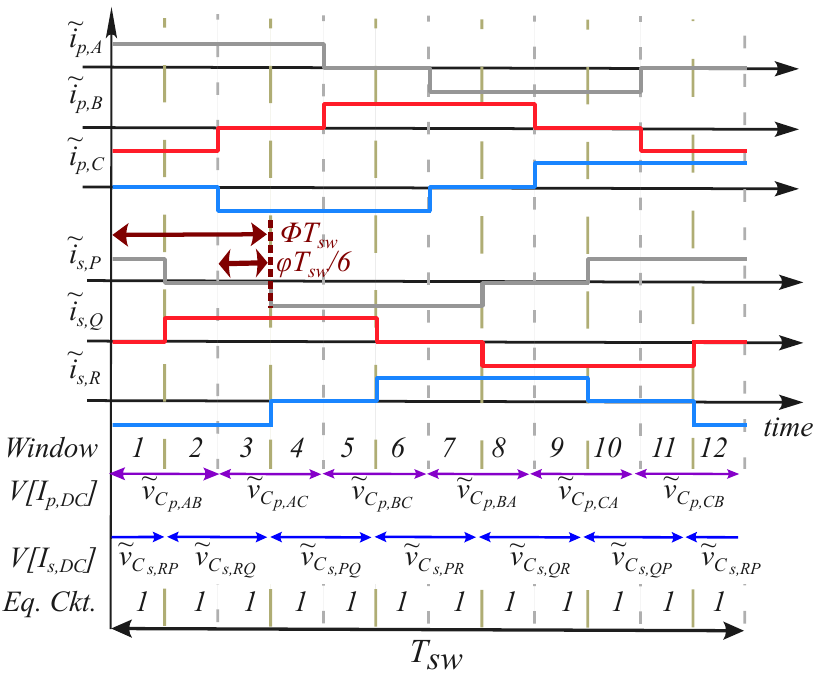}}
\caption{Selected analytical waveforms of the capacitive bridge transformer topology depicting the equivalent circuit mapping when $1/6<\Phi<1/3$.}
\label{fig:waveforms_3Phase2}
\end{figure}

Next, let us consider the case when $1/6<\Phi<1/3$. Fig.~\ref{fig:waveforms_3Phase2} qualitatively illustrates the throw currents of the two CSI bridges for the range of phase-shifts under consideration. It may be observed that during the entire window of operation, the ideal equivalent three-phase circuit can simply be represented using one equivalent circuit which is illustrated in Fig.~\ref{fig:eq_ckt1}. While this is in contrast to Case 1, (\ref{eq:A_m1})-(\ref{eq:A_i4}) are still applicable. A similar methodology adopted in Section \ref{sec:case1} can be applied to calculate the power flow equation for Case 2. As noted in the figure, symbol $\varphi$ still represents the per-unitized phase-shift for 1/6 cycle and $\varphi=6\Phi-1$ for $1/6<\Phi<1/3$. In other words, varying $\varphi$ from 0 to 1 spans 1/6 of the switching cycle.

Considering Fig.~\ref{fig:waveforms_3Phase2}, as a precursor to deriving the averaged voltage across the delta capacitor, $C_{p,AB}$, the steady-state capacitor voltages can be derived as noted in (\ref{eq:A_vmean2})-(\ref{eq:A_vmean3})   
%
%
%
%
where, the capacitor current expressions have been summarized in Table \ref{tab:map_iCAB}. Using (\ref{eq:A_vmean1})-(\ref{eq:A_vmean3}), $\overline{v}_{C_{p,AB}}|_{T_{sw}/6}$ can be derived as characterized in (\ref{eq:B_vmean4}).
\begin{multline}
    \overline{v}_{C_{p,AB}}|_{T_{sw}/6} = V_0 + \\ \frac{T_{sw}}{36C_p} \left[ \frac{3/C_m+1/C_s}{1/C_{eq}}2\tilde{i}_p + \frac{1/C_s}{1/C_{eq}}\tilde{i}_s(1-4\varphi+2\varphi^2)\right] \label{eq:B_vmean4}
\end{multline}

\noindent where, similar to Case 1, the initial steady-state capacitor voltage, $V_0$, can be calculated by assuming that the voltage averaged over the full fundamental cycle, $T_{sw}$, is zero. Using the Table \ref{tab:map_iCAB} current map, the current expressions for Window 1-2 can be substituted into (\ref{eq:A_Vinit1}) to arrive at (\ref{eq:B_Vinit2}).
\begin{align}
V_0 = -\frac{T_{sw}}{18}\frac{C_{eq}}{C_p}[(3/C_m+1/C_s)\tilde{i}_p+(2-\varphi )\tilde{i}_s/C_s]\label{eq:B_Vinit2}
\end{align}
Using (\ref{eq:B_vmean4}), (\ref{eq:B_Vinit2}), the power flow expression can be derived to be as noted in (\ref{eq:B_P1}) or (\ref{eq:B_P2}) in terms of the per-unitized phase-shift $\varphi$ or complete cycle phase-shift $\Phi$ respectively.
\begin{multline}
    P_p (0<\varphi<1) = \\ - \frac{T_{sw}}{36C_pC_s}\frac{I_{p,DC}I_{s,DC}}{3/C_m+1/C_p+1/C_s}(3+2\varphi-2\varphi^2) \label{eq:B_P1}
\end{multline}
\begin{multline}
    P_p (1/6<\Phi<1/3) = \\ - \frac{T_{sw}}{36C_pC_s}\frac{I_{p,DC}I_{s,DC}}{3/C_m+1/C_p+1/C_s}(-1+36\Phi-72\Phi^2) \label{eq:B_P2}
\end{multline}

\color{black}
Similar to Case 1 analysis, assuming lossless converter operation, the power flow expression can be used to derive the voltage and current transfer ratio expressions as noted in (\ref{eq:B_ratio}).
\begin{multline}
    1/6<\Phi<1/3: \frac{V_{s,DC}}{V_{p,DC}} = \frac{I_{p,DC}}{I_{s,DC}} = \\ -\frac{36C_pC_sR_s}{T_{sw}}\frac{3/C_m+1/C_p+1/C_s}{(-1+36\Phi-72\Phi^2)} \label{eq:B_ratio}
\end{multline} 
%

Lastly, the apparent power expression for Case 2 adopts the same format of (\ref{eq:S2_Case1}) from Case 1 but with the weighting parameters, $s_1$ and $s_2$, as noted in (\ref{eq:S2_Case2})-(\ref{eq:S3_Case2}).

\begin{align}
    s_1 &= 8\Phi^3-6\Phi^2+\frac{4\Phi}{3}-\frac{1}{12} \label{eq:S2_Case2}\\
    s_2 &= -3\Phi^3+\frac{3\Phi^2}{2}-\frac{11}{108} \label{eq:S3_Case2}
\end{align}

\color{black}

\subsection{Case 3: $1/3<\Phi<1/2$}\label{sec:case3}

\begin{figure}[htbp]
\centerline{\includegraphics[width=0.9\columnwidth]{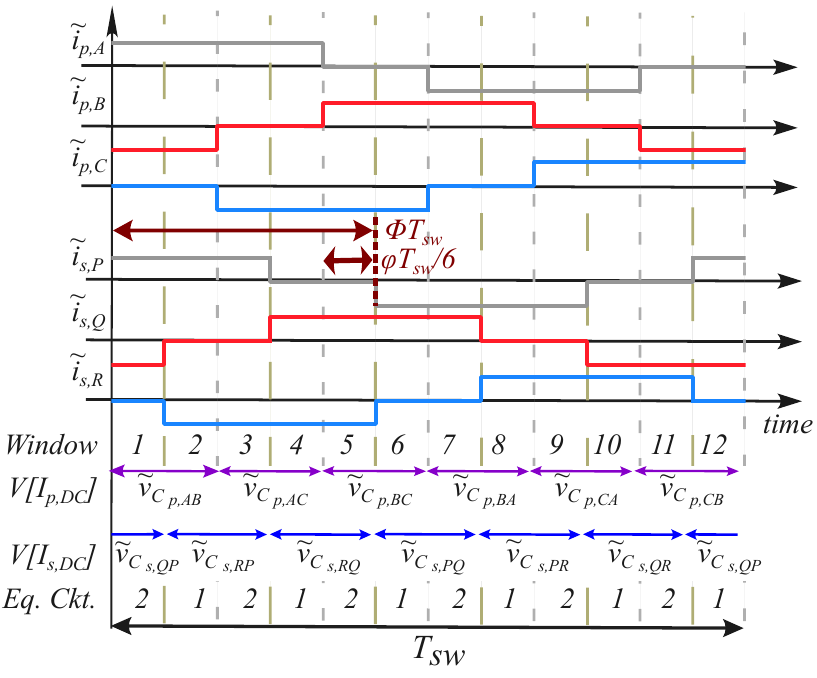}}
\caption{Selected analytical waveforms of the capacitive bridge transformer topology depicting the equivalent circuit mapping when $1/6<\Phi<1/3$.}
\label{fig:waveforms_3Phase3}
\end{figure}

Lastly, let us consider the case when $1/3<\Phi<1/2$. Fig.~\ref{fig:waveforms_3Phase3} qualitatively illustrates the throw currents of the two CSI bridges for the range of phase-shifts under consideration. In Case 3, during Interval 1, 3, \& 5 of operation, the ideal equivalent three-phase circuit can be represented as shown in Fig.~\ref{fig:eq_ckt2}. However, during Interval 2, 4, \& 6 of operation, the ideal equivalent three-phase circuit can be represented as shown in Fig.~\ref{fig:eq_ckt1}. As can be predicted using symmetry, while the same equivalent circuits as in Case 1 appear in this case, the sequencing has reversed. Hence, (\ref{eq:A_m1})-(\ref{eq:A_i5}) are applicable for the considered window of phase-shifts. A similar methodology adopted in Section \ref{sec:case1} and Section \ref{sec:case2} can be applied to calculate the power flow equation for $1/3<\Phi<1/2$. As noted in the figure, symbol $\varphi$ still represents the per-unitized phase-shift for 1/6 cycle and $\varphi=6\Phi-2$ for $1/3<\Phi<1/2$. In other words, varying $\varphi$ from 0 to 1 spans 1/6 of the switching cycle. 

Considering Fig.~\ref{fig:waveforms_3Phase3}, as a precursor to deriving the averaged voltage across the delta capacitor, $C_{p,AB}$, the steady-state capacitor voltages can be derived as noted in (\ref{eq:A_vmean2})-(\ref{eq:A_vmean3}).   
%
%
%
%
where, the capacitor current expressions have been summarized in Table \ref{tab:map_iCAB}. Using (\ref{eq:A_vmean1})-(\ref{eq:A_vmean3}), $\overline{v}_{C_{p,AB}}|_{T_{sw}/6}$ can be derived as characterized in (\ref{eq:C_vmean4}).
\begin{multline}
    \overline{v}_{C_{p,AB}}|_{T_{sw}/6} = V_0 + \\ \frac{T_{sw}}{36C_p} \left[ \frac{3/C_m+1/C_s}{1/C_{eq}}2\tilde{i}_p + \frac{1/C_s}{1/C_{eq}}\tilde{i}_s(-1-2\varphi+\varphi^2)\right] \label{eq:C_vmean4}
\end{multline}

\noindent where, similar to Case 1, the initial steady-state capacitor voltage, $V_0$, can be calculated by assuming that the voltage averaged over the full fundamental cycle, $T_{sw}$, is zero. Using the Table \ref{tab:map_iCAB} current map, the current expressions for Window 1-2 can be substituted into (\ref{eq:A_Vinit1}) to arrive at (\ref{eq:C_Vinit2}).
\begin{align}
V_0 = -\frac{T_{sw}}{18}\frac{C_{eq}}{C_p}[(3/C_m+1/C_s)\tilde{i}_p+(1-2\varphi)\tilde{i}_s/C_s]\label{eq:C_Vinit2}
\end{align}
Using (\ref{eq:C_vmean4}), (\ref{eq:C_Vinit2}), the power flow expression can be derived to be as noted in (\ref{eq:C_P1}) or (\ref{eq:C_P2}) in terms of the per-unitized phase-shift $\varphi$ or complete cycle phase-shift $\Phi$ respectively.
\begin{multline}
    P_p (0<\varphi<1) = \\ -\frac{T_{sw}}{36C_pC_s}\frac{I_{p,DC}I_{s,DC}}{3/C_m+1/C_p+1/C_s}(3-2\varphi-\varphi^2) \label{eq:C_P1}
\end{multline}
\begin{multline}
    P_p (1/3<\Phi<1/2) = \\ -\frac{T_{sw}}{12C_pC_s}\frac{I_{p,DC}I_{s,DC}}{3/C_m+1/C_p+1/C_s}(1+4\Phi-12\Phi^2) \label{eq:C_P2}
\end{multline}
\color{black}
Further, assuming an lossless converter operation, the power flow expression can be used to derive the voltage and current transfer ratio expressions as noted in (\ref{eq:C_ratio}). 
\begin{multline}
    1/3<\Phi<1/2: \frac{V_{s,DC}}{V_{p,DC}} = \frac{I_{p,DC}}{I_{s,DC}} = \\ -\frac{12C_pC_sR_s}{T_{sw}}\frac{3/C_m+1/C_p+1/C_s}{(1+4\Phi-12\Phi^2)} \label{eq:C_ratio}
\end{multline}
\color{black}

\color{black}

Lastly, using (\ref{eq:S}), (\ref{eq:rms1}), (\ref{eq:rms2}), the apparent power expression can also be derived as noted in (\ref{eq:S2_Case1}) for the Case 1 scenario. However, the weighting parameters, $s_1$ and $s_2$, for Case 3 are noted in (\ref{eq:S2_Case3})-(\ref{eq:S3_Case3}).

\begin{align}
    s_1 &= 4\Phi^3-5\Phi^2+2\Phi-\frac{29}{108} \label{eq:S2_Case3}
    \\
    s_2 &= 6\Phi^3-6\Phi^2+\frac{3\Phi}{2}+\frac{1}{108} \label{eq:S3_Case3}
\end{align}

Next, Section \ref{sec:experiments} presents results from an experimental prototype validating the analytical models for the real and apparent power calculations as well as voltage and current transfer ratios as a function of phase-shift.

\color{black}

\section{Experimental Results} \label{sec:experiments}

\begin{figure}[htbp]
\centerline{\includegraphics[width=0.9\columnwidth]{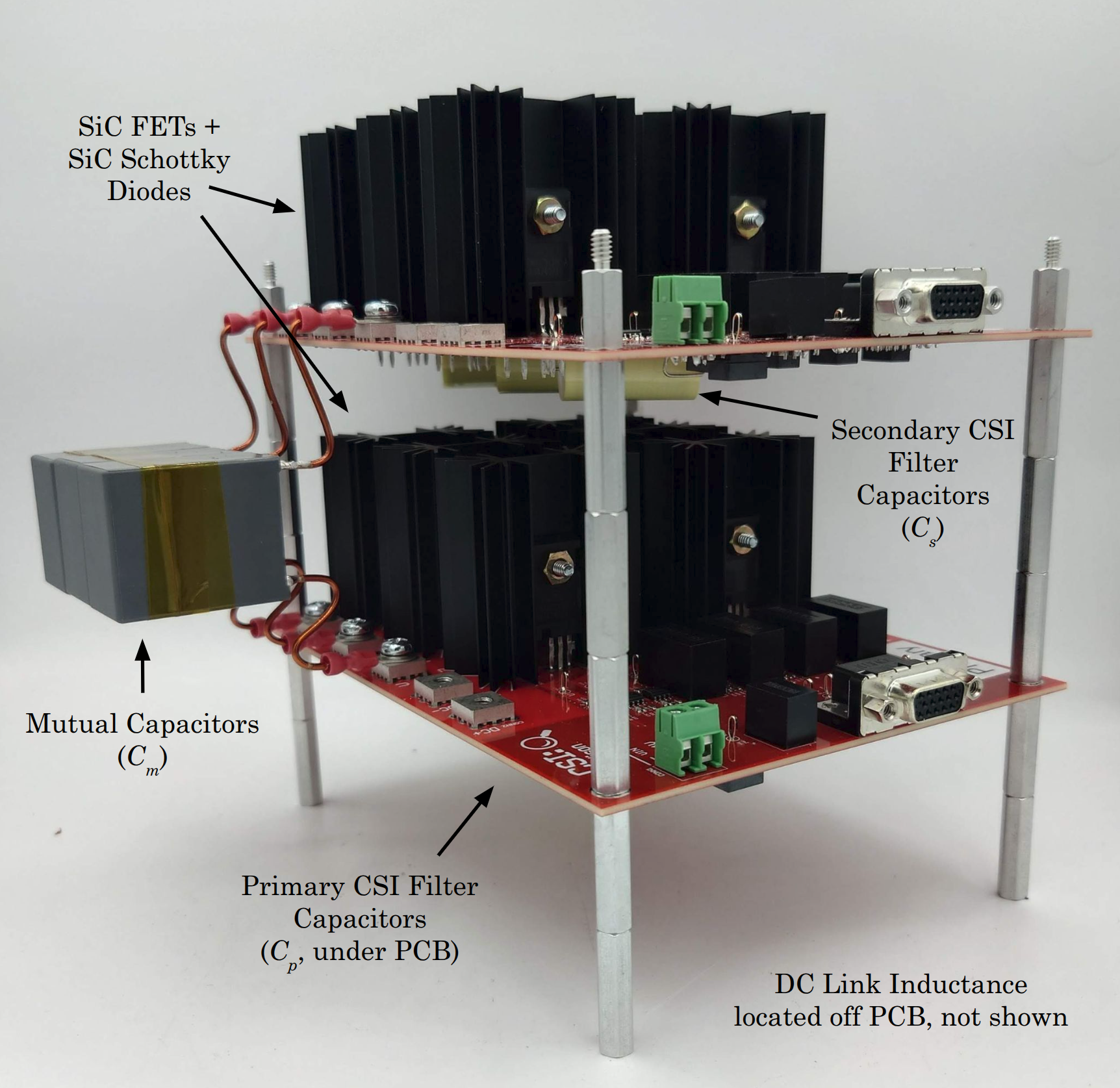}}
\captionsetup{font={color=black, footnotesize}, belowskip=-4pt}
\caption{The prototype Capacitive Bridge Transformer.}
\label{fig:cbt-proto}
\end{figure}

\begin{figure}[htbp]
\centering
\begin{tikzpicture}[xscale=0.15, yscale=0.5]

\definecolor{phaseBlue}{HTML}{0071BD}
\definecolor{phaseBlueDark}{HTML}{01538a}

\definecolor{phaseOrange}{HTML}{D95319}
\definecolor{phaseOrangeDark}{HTML}{b54b1d}

\definecolor{phaseYellow}{HTML}{EAB120}
\definecolor{phaseYellowDark}{HTML}{c99a20}

\definecolor{phiArrow}{HTML}{D3D3D3}


\foreach \x/\label in {0/0,8.33/$\frac{1}{6} T_{sw}$,16.67/$\frac{2}{6} T_{sw}$, 25/$\frac{3}{6} T_{sw}$, 33.33/$\frac{4}{6} T_{sw}$, 41.67/$\frac{5}{6} T_{sw}$, 50/$T_{sw}$} {
  \draw (\x,0.15) -- (\x,-0.15) node[below]{\label};
}

\def\pre{150/100}
\def\post{250/100}
\def\phaseperiod{1667/100}
\def\complement{2500/100}

\def\AStartHigh{0}
\def\AEndHigh{\phaseperiod}
\def\BStartHigh{\phaseperiod}
\def\BEndHigh{2*\phaseperiod}
\def\CStartHigh{2*\phaseperiod}
\def\CEndHigh{50}

\def\AStartLow{\complement}
\def\AEndLow{\complement + \phaseperiod}
\def\BStartEndLow{\complement + \phaseperiod}
\def\BEndEndLow{50}
\def\BStartInitLow{0}
\def\BEndInitLow{\complement - \phaseperiod}
\def\CStartLow{\complement - \phaseperiod}
\def\CEndLow{\complement}

\def\PHBA{13}
\def\PHBB{12}
\def\PHBC{11}

\def\PLBA{10}
\def\PLBB{9}
\def\PLBC{8}

\newcommand{\baseline}[1]{\draw[gray!40] (0,#1) -- (50,#1);}

\newcommand{\pulse}[5]{%
  \pgfmathsetmacro{\pulseStart}{#1}%
  \pgfmathsetmacro{\pulseEnd}{#2}%

  \ifnum #5=1
    \draw[#4,very thick]
      (\pulseStart,#3) -- (\pulseStart,#3+0.6) --
      (\pulseEnd,#3+0.6) -- (\pulseEnd,#3);
  \else
    \ifdim\pulseStart pt=0pt
      \ifdim\pulseEnd pt=50pt
        \draw[#4,very thick]
          (0,#3+0.6) -- (50,#3+0.6);
      \else
        \draw[#4,very thick]
          (0,#3+0.6) --
          (\pulseEnd,#3+0.6) -- (\pulseEnd,#3);
      \fi
    \else
      \ifdim\pulseEnd pt=50pt
        \draw[#4,very thick]
          (\pulseStart,#3) -- (\pulseStart,#3+0.6) --
          (50,#3+0.6);
      \else
        \draw[#4,very thick]
          (\pulseStart,#3) -- (\pulseStart,#3+0.6) --
          (\pulseEnd,#3+0.6) -- (\pulseEnd,#3);
      \fi
    \fi
  \fi
}

\baseline{\PHBA}
\baseline{\PHBB}
\baseline{\PHBC}

\baseline{\PLBA}
\baseline{\PLBB}
\baseline{\PLBC}

\pulse{\AStartHigh}{\AEndHigh+\post}{\PHBA}{phaseBlue}{1}
\pulse{\BStartHigh}{\BEndHigh+\post}{\PHBB}{phaseOrange}{1}
\pulse{\AStartHigh}{\AStartHigh+\post}{\PHBC}{phaseYellow}{0}
\pulse{\CStartHigh}{\CEndHigh}{\PHBC}{phaseYellow}{0}

\pulse{\AStartLow}{\AEndLow+\post}{\PLBA}{phaseBlue}{1}
\pulse{\BStartInitLow}{\BEndInitLow+\post}{\PLBB}{phaseOrange}{0}
\pulse{\BStartEndLow}{\BEndEndLow}{\PLBB}{phaseOrange}{0}
\pulse{\CStartLow}{\CEndLow+\post}{\PLBC}{phaseYellow}{1}

\node[left] at (0,\PHBA+0.3) {AH};
\node[left] at (0,\PHBB+0.3) {BH};
\node[left] at (0,\PHBC+0.3) {CH};

\node[left] at (0,\PLBA+0.3) {AL};
\node[left] at (0,\PLBB+0.3) {BL};
\node[left] at (0,\PLBC+0.3) {CL};

\draw[dashed,gray] (0*\phaseperiod,14) -- (0*\phaseperiod,0);
\draw[dashed,gray] (0.5*\phaseperiod,14) -- (0.5*\phaseperiod,0);
\draw[dashed,gray] (1*\phaseperiod,14) -- (1*\phaseperiod,0);
\draw[dashed,gray] (1.5*\phaseperiod,14) -- (1.5*\phaseperiod,0);
\draw[dashed,gray] (2*\phaseperiod,14) -- (2*\phaseperiod,0);
\draw[dashed,gray] (2.5*\phaseperiod,14) -- (2.5*\phaseperiod,0);
\draw[dashed,gray] (3*\phaseperiod,14) -- (3*\phaseperiod,0);


\def\pre{150/100}
\def\post{250/100}
\def\delay{1250/100}
\def\phaseperiod{1667/100}
\def\complement{2500/100}

\def\AStartHigh{\delay}
\def\AEndHigh{\delay + \phaseperiod}
\def\BStartHigh{\delay + \phaseperiod}
\def\BEndHigh{50 - \phaseperiod + \delay}
\def\CStartEndHigh{50 - \phaseperiod + \delay}
\def\CEndEndHigh{50}
\def\CStartInitHigh{0}
\def\CEndInitHigh{\delay}

\def\AStartEndLow{\delay + \complement}
\def\AEndEndLow{50}
\def\AStartInitLow{0}
\def\AEndInitLow{\complement - 2*\phaseperiod + \delay}
\def\BStartLow{\complement - 2*\phaseperiod + \delay}
\def\BEndLow{0.5*\phaseperiod + \delay}
\def\CStartLow{0.5*\phaseperiod + \delay}
\def\CEndLow{3/2*\phaseperiod + \delay}

\def\SHTA{5}
\def\SHTB{4}
\def\SHTC{3}

\def\SLTA{2}
\def\SLTB{1}
\def\SLTC{0}

\baseline{\SHTA}
\baseline{\SHTB}
\baseline{\SHTC}
\baseline{\SLTA}
\baseline{\SLTB}
\baseline{\SLTC}

\pulse{\AStartHigh}{\AEndHigh+\post}{\SHTA}{phaseBlueDark}{1}
\draw[<-> , ultra thick, phiArrow]
  (\AStartHigh,\SHTA+0.4) -- (0,\SHTA+0.4)
  node[midway, above, black]{$\Phi$};

\pulse{\BStartHigh}{\BEndHigh+\post}{\SHTB}{phaseOrangeDark}{1}
\draw[<-> , ultra thick, phiArrow]
  (\BStartHigh,\SHTB+0.4) -- (\phaseperiod,\SHTB+0.4)
  node[midway, above, black]{$\Phi$};

\pulse{\CStartInitHigh}{\CEndInitHigh+\post}{\SHTC}{phaseYellowDark}{0}
\pulse{\CStartEndHigh}{\CEndEndHigh}{\SHTC}{phaseYellowDark}{0}
\draw[<-> , ultra thick, phiArrow]
  (\CStartEndHigh,\SHTC+0.4) -- (2*\phaseperiod,\SHTC+0.4)
  node[midway, above, black]{$\Phi$};

\pulse{\AStartInitLow}{\AEndInitLow+\post}{\SLTA}{phaseBlueDark}{0}
\pulse{\AStartEndLow}{\AEndEndLow}{\SLTA}{phaseBlueDark}{0}
\draw[<-> , ultra thick, phiArrow]
  (\AStartEndLow,\SLTA+0.4) -- (3/2*\phaseperiod,\SLTA+0.4)
  node[midway, above, black]{$\Phi$};

\pulse{\BStartLow}{\BEndLow+\post}{\SLTB}{phaseOrangeDark}{1}
\draw[<- , ultra thick, phiArrow]
  (\BStartLow,\SLTB+0.4) -- (0,\SLTB+0.4)
  node[midway, above, black]{};
\draw[-> , ultra thick, phiArrow]
  (50,\SLTB+0.4) -- (5/2*\phaseperiod,\SLTB+0.4)
  node[midway, above, black]{$\Phi$};

\pulse{\CStartLow}{\CEndLow+\post}{\SLTC}{phaseYellowDark}{1}
\draw[<-> , ultra thick, phiArrow]
  (\CStartLow,\SLTC+0.4) -- (1/2*\phaseperiod,\SLTC+0.4)
  node[midway, above, black]{$\Phi$};

\node[left] at (0,\SHTA+0.3) {PH};
\node[left] at (0,\SHTB+0.3) {QH};
\node[left] at (0,\SHTC+0.3) {RH};

\node[left] at (0,\SLTA+0.3) {PL};
\node[left] at (0,\SLTB+0.3) {QL};
\node[left] at (0,\SLTC+0.3) {RL};


\end{tikzpicture}
\captionsetup{font={color=black, footnotesize}, belowskip=-4pt}
\caption{Six-step timing diagram of the prototype Capacitive Bridge Transformer.}
\label{fig:cbt-proto-timing}
\end{figure}
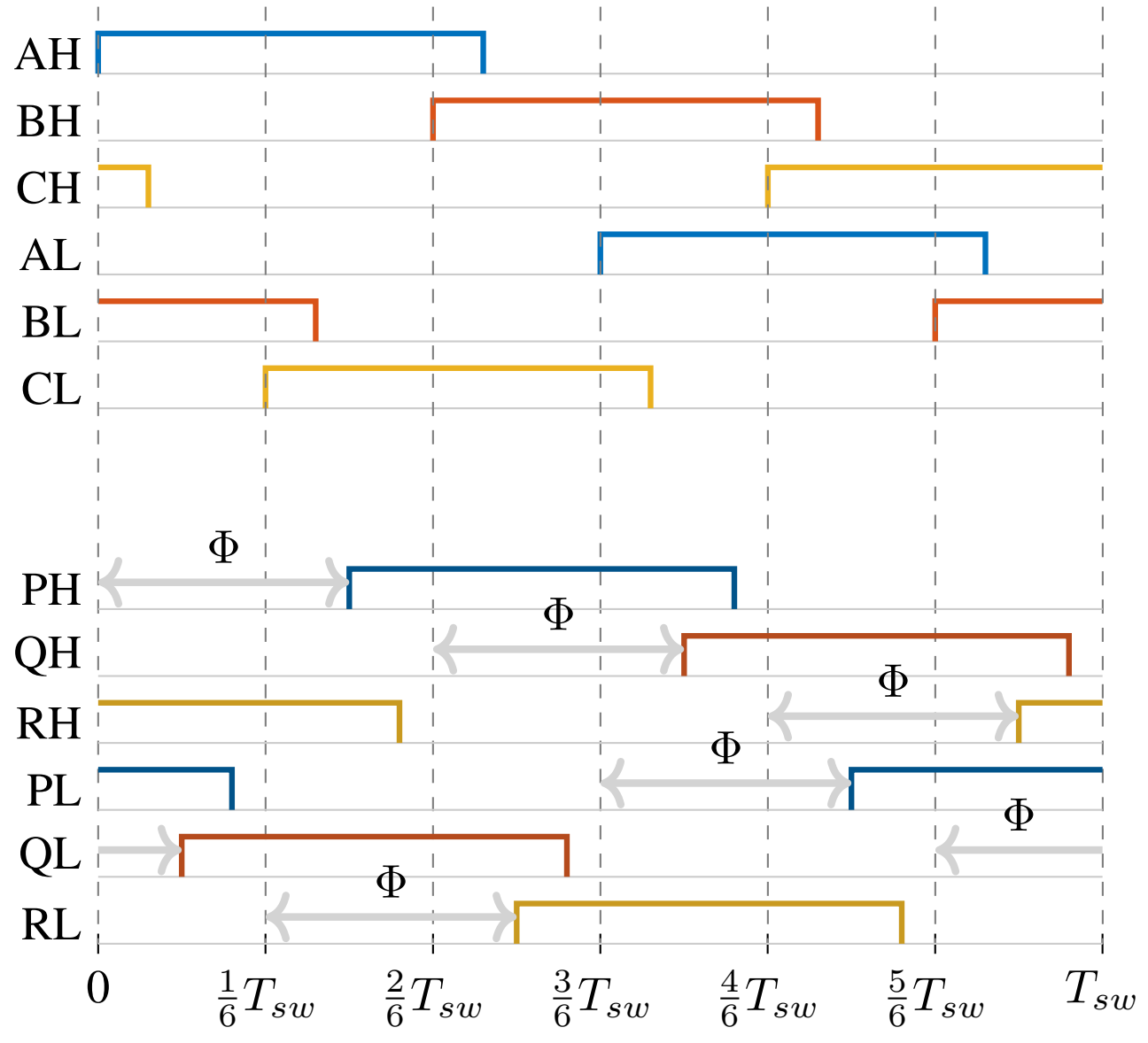

A bench-scale CBT prototype shown in Fig. \ref{fig:cbt-proto} was built to verify the analytical model presented in Sections \ref{sec:case1} through \ref{sec:case3}. The circuit diagram is shown in Fig. \ref{fig:cbt-proto-circuit} with the values of passives in the caption. The CBT consists of two separate three-phase current source inverters with mutual capacitors linking the inverters together at their AC terminals \cite{Nowakowski2025CBT}. Although monolithic bidirectional blocking GaN devices would be ideal candidates for prototype realization, limited availability prevented their use. Instead, the switching devices are discrete TO-247 package, 1200 V, 20 A SiC FETs with SiC Schottkey diodes of equal rating for reverse blocking capabilities. During experiments, the input of the primary CSI was fed by a DC power supply operating in constant-current mode. The DC link of \color{black} the \color{black} secondary CSI was connected to a resistive bank to serve as the load. The CSI nominal 6-step switching frequency was 20 kHz, \color {black} shown by the timing diagram in Fig. \ref{fig:cbt-proto-timing}\color{black}.

\begin{figure}[htbp]
\centering
\trimbox{0.35cm 0cm 0.35cm 0cm}{ 
\scalebox{0.35}{
\begin{circuitikz}[american,voltage dir=old,line width=1pt]
\ctikzset{diodes/scale=0.7}
\ctikzset{bipoles/thickness=1}
\ctikzset{bipole label style/.style={font=\LARGE}}
\ctikzset{bipole annotation style/.style={font=\LARGE}}
\draw           
        (0,0)   coordinate(start)
                to [short]++(0,2) to [L,l_=$\frac{L_p}{2}$] ++ (0,3)
                to [V,v=$ $,l_=$V_{p,DC}$,i_>=$I_{p,DC}$]++(0,3)
                to [L,l_=$\frac{L_p}{2}$] ++ (0,3) to [short]++(0,2)
                to [short, -*] ++ (4,0) coordinate(a1)
                to [short, -*] ++ (1.5,0) coordinate(b1)
                to [short] ++ (1.5,0) coordinate(c1)
                to [diode] ++ (0,-1)
                node[nigfete, bodydiode, anchor=D] (CH) {} ++ (0,-1.5)
                to [short, -*] ++ (0,-7.5) coordinate (cp) 
                to [short] ++ (0,-0.5)
                to [diode] ++ (0,-1)
                node[nigfete, bodydiode, anchor=D] (CL) {} ++ (0,-1.5)
                to [short, -*] ++ (-1.5, 0)
                to [short, -*] ++ (-1.5, 0)
                to [short] ++ (-4, 0) 
        (b1)    to [diode] ++ (0,-1)
                node[nigfete, bodydiode, anchor=D] (BH) {} ++ (0,-1.5)
                to [short, -*] ++ (0,-4) coordinate (bp) 
                to [short] ++ (0,-4)
                to [diode] ++ (0,-1)
                node[nigfete, bodydiode, anchor=D] (BL) {} ++ (0,-1.5)
        (a1)    to [diode] ++ (0,-1)
                node[nigfete, bodydiode, anchor=D] (AH) {} ++ (0,-1.5)
                to [short, -*] ++ (0,-0.5) coordinate (ap) 
                to [short] ++ (0,-7.5)
                to [diode] ++ (0,-1)
                node[nigfete, bodydiode, anchor=D] (AL) {} ++ (0,-1.5)

        (ap)    to [short]++(0.75,0) to [crossing]++(1.5,0) to [crossing]++(1.5,0)
                to [short]++(1.25,0) coordinate(f1) to [short]++(3,0)
                to [C,l=$C_m$]++(2,0)
                to [short]++(3,0) coordinate(f2) to [short]++(1.25,0)
                to [short]++(0.75,0)
                
        (bp)    to [short]++(0.75,0) to [crossing]++(1.5,0)
                to [short]++(1.25,0) to [short]++(1,0) to [crossing]++(2,0)
                to [C,l=$C_m$]++(2,0)
                to [crossing]++(2,0) to [short]++(1,0) to [short]++(1.25,0)
                to [crossing]++(1.5,0) to [short]++(0.75,0)
                
        (cp)    to [short]++(0.75,0)
                to [short]++(1.25,0) to [short]++(3,0)
                to [C,l=$C_m$]++(2,0) 
                to [short]++(3,0) to [short]++(1.25,0)
                to [crossing]++(1.5,0) to [crossing]++(1.5,0) to [short]++(0.75,0)

        (f1)    to [C,l_=$C_p$,*-*]++(0,-3.5)
                to [C,l_=$C_p$,-*]++(0,-3.5)
                ++(2,0) to [C,l=$C_p$,*-]++(0,3.5) to [short,-*]++(0,3.5)

        (f2)    to [C,l=$C_s$,*-*]++(0,-3.5)
                to [C,l=$C_s$,-*]++(0,-3.5)
                ++(-2,0) to [C,l_=$C_s$,*-]++(0,3.5) to [short,-*]++(0,3.5)

        (24,0) coordinate (start2)
                to [short]++(0,2) to [L,l=$\frac{L_s}{2}$,mirror]++(0,3)
                to [R,v>=$V_{s,DC}$,l=$R_s$,i^<=$I_{s,DC}$]++(0,3)
                to [L,l=$\frac{L_s}{2}$,mirror]++(0,3) to [short]++(0,2)
                to [short, -*] ++ (-2,0) coordinate(c2)
                to [short, -*] ++ (-1.5,0) coordinate(b2)
                to [short] ++ (-1.5,0) coordinate(a2)
                node[nigfete, bodydiode, anchor=S, yscale=-1] (PH) {} ++ (0,-1.5)
                to [diode,invert] ++ (0,-1)
                to [short, -*] ++ (0,-0.5) coordinate (up) 
                to [short] ++ (0,-7.5)
                node[nigfete, bodydiode, anchor=S, yscale=-1] (PL) {} ++ (0,-1.5)
                to [diode,invert] ++ (0,-1)
                to [short, -*] ++ (1.5, 0)
                to [short, -*] ++ (1.5, 0)
                to [short] ++ (2, 0)                
        (b2)    node[nigfete, bodydiode, anchor=S, yscale=-1] (QH) {} ++ (0,-1.5)
                to [diode,invert] ++ (0,-1)
                to [short, -*] ++ (0,-4) coordinate (vp) 
                to [short] ++ (0,-4)
                node[nigfete, bodydiode, anchor=S, yscale=-1] (QL) {} ++ (0,-1.5)
                to [diode,invert] ++ (0,-1)
        (c2)    node[nigfete, bodydiode, anchor=S, yscale=-1] (RH) {} ++ (0,-1.5)
                to [diode,invert] ++ (0,-1)
                to [short, -*] ++ (0,-7.5) coordinate (wp) 
                to [short] ++ (0,-0.5)
                node[nigfete, bodydiode, anchor=S, yscale=-1] (RL) {} ++ (0,-1.5)
                to [diode,invert] ++ (0,-1)
        ;

        \draw (AH.G) ++ (0.2,0) node[below]{AH};
        \draw (AL.G) ++ (0.2,0) node[below]{AL};

        \draw (BH.G) ++ (0.2,0) node[below]{BH};
        \draw (BL.G) ++ (0.2,0) node[below]{BL};
        
        \draw (CH.G) ++ (0.2,0) node[below]{CH};
        \draw (CL.G) ++ (0.2,0) node[below]{CL};

        \draw (PH.G) ++ (0.2,0) node[above]{PH};
        \draw (PL.G) ++ (0.2,0) node[above]{PL};

        \draw (QH.G) ++ (0.2,-0.05) node[above]{QH};
        \draw (QL.G) ++ (0.2,-0.05) node[above]{QL};
        
        \draw (RH.G) ++ (0.2,0) node[above]{RH};
        \draw (RL.G) ++ (0.2,0) node[above]{RL};

        \draw[dashed, line width=1.5pt] (2.5,-0.5) -- (2.5,13.5);
        \draw[-{Triangle}, line width=2pt] (1.5,2) |- (3.5,5);
        
\end{circuitikz}
}
}
\captionsetup{font={color=black, footnotesize}, belowskip=-4pt}
\caption{Circuit diagram of the prototype Capacitive Bridge Transformer. $C_p$, $C_s$ = 0.1 $\mu$F. $C_m$ = 5.0 $\mu$F. $L_p$, $L_s$ = 940 $\mu$H. $R_s$ = 20 $\Omega$. The "looking into" arrow shows where apparent and reactive power measurements were taken.}
\label{fig:cbt-proto-circuit}
\end{figure}
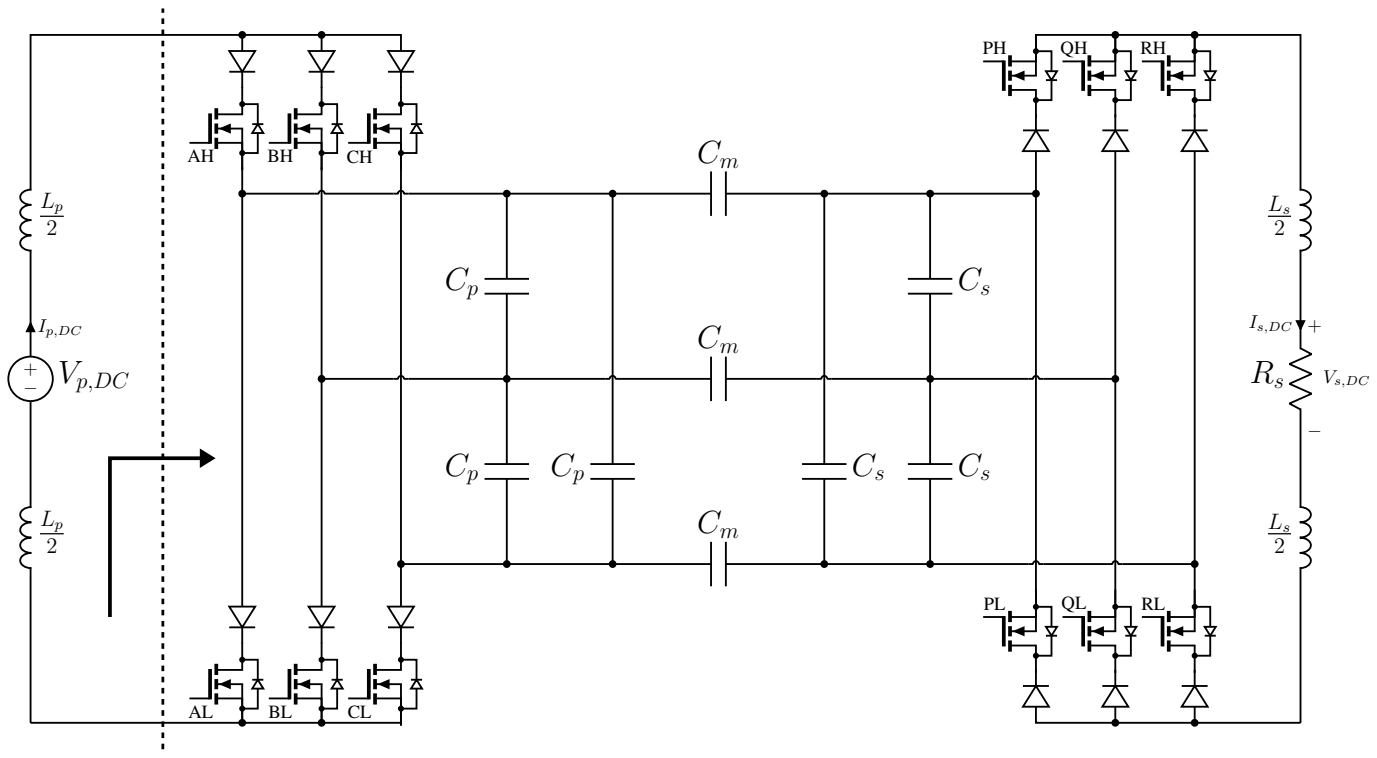

\begin{figure}[htbp]
\captionsetup{font={color=black, footnotesize}, belowskip=-4pt}
\centerline{\includegraphics[width=0.95\columnwidth]{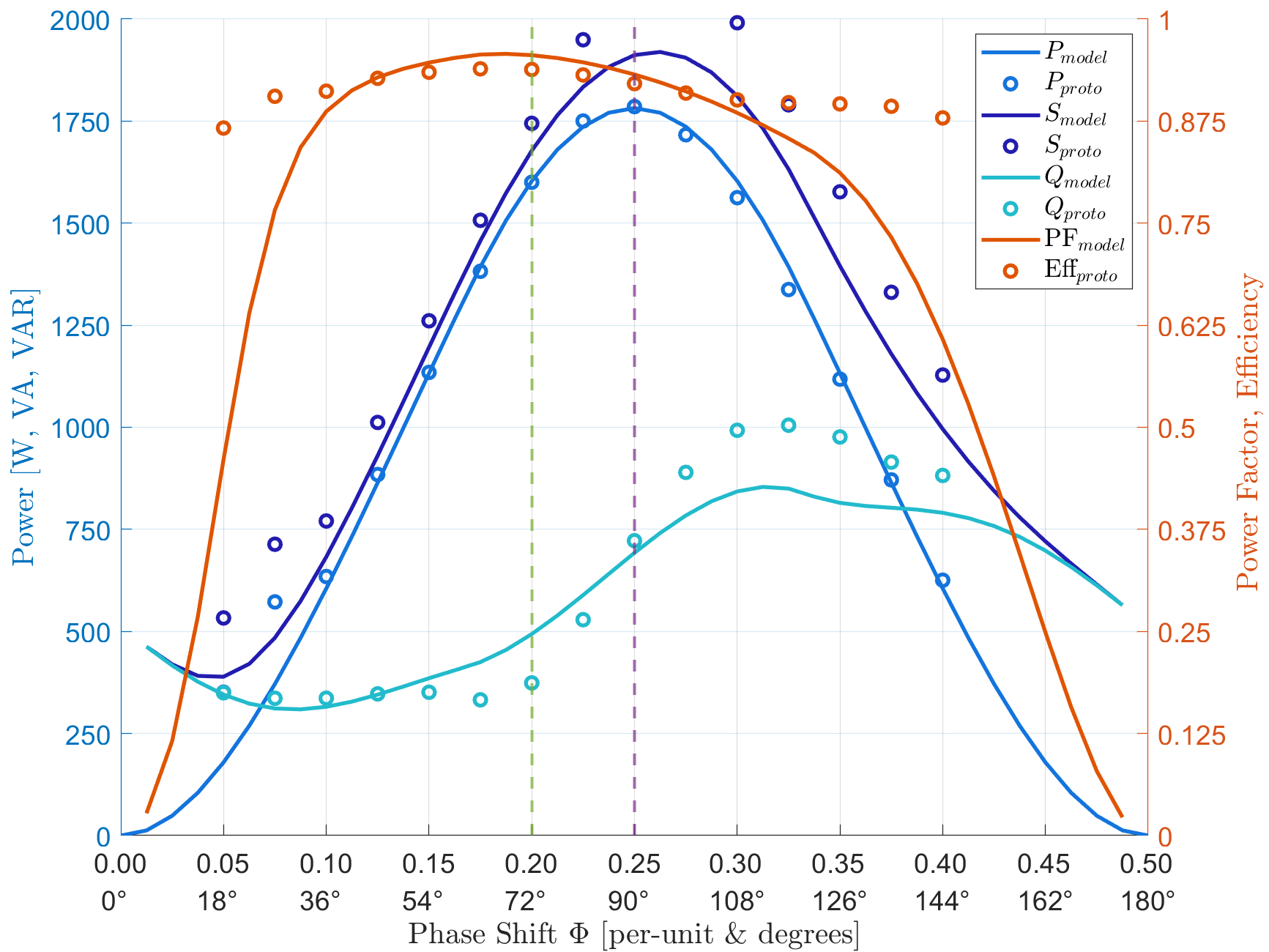}}
\caption{Output Power predicted by analytical model vs measured on prototype, 3-$\phi$ $\Delta$-connected capacitors.}
\label{fig:power-comparison}
\end{figure}

\color{black} To experimentally verify the topology and analytical model, the first round of experiments targeted an effective turns ratio of 1 at peak power flow. Eq. \eqref{eq:B_ratio} was used to select the load resistance resulting in unity gain at a phase shift of 90 degrees, $R_s=20$ $\Omega$. The measured DC-DC power of the prototype versus primary-to-secondary phase shift $\Phi$ for a constant primary DC link current of 8 A is plotted in Fig. \ref{fig:power-comparison}, along with the internal apparent power, reactive power, and power factor predicted by the analytical model. For this DC-DC CBT, the internal AC power quantities were measured between the primary link inductors and switching bridge as shown by the ``looking into" arrow in Fig. \ref{fig:cbt-proto-circuit}. This switching bridge internal apparent power is the result of the DC link current and AC voltage that appears across the switching devices.

\begin{figure}[htbp]
\captionsetup{font={color=black, footnotesize}, belowskip=-4pt}
\centerline{\includegraphics[width=0.9\columnwidth]{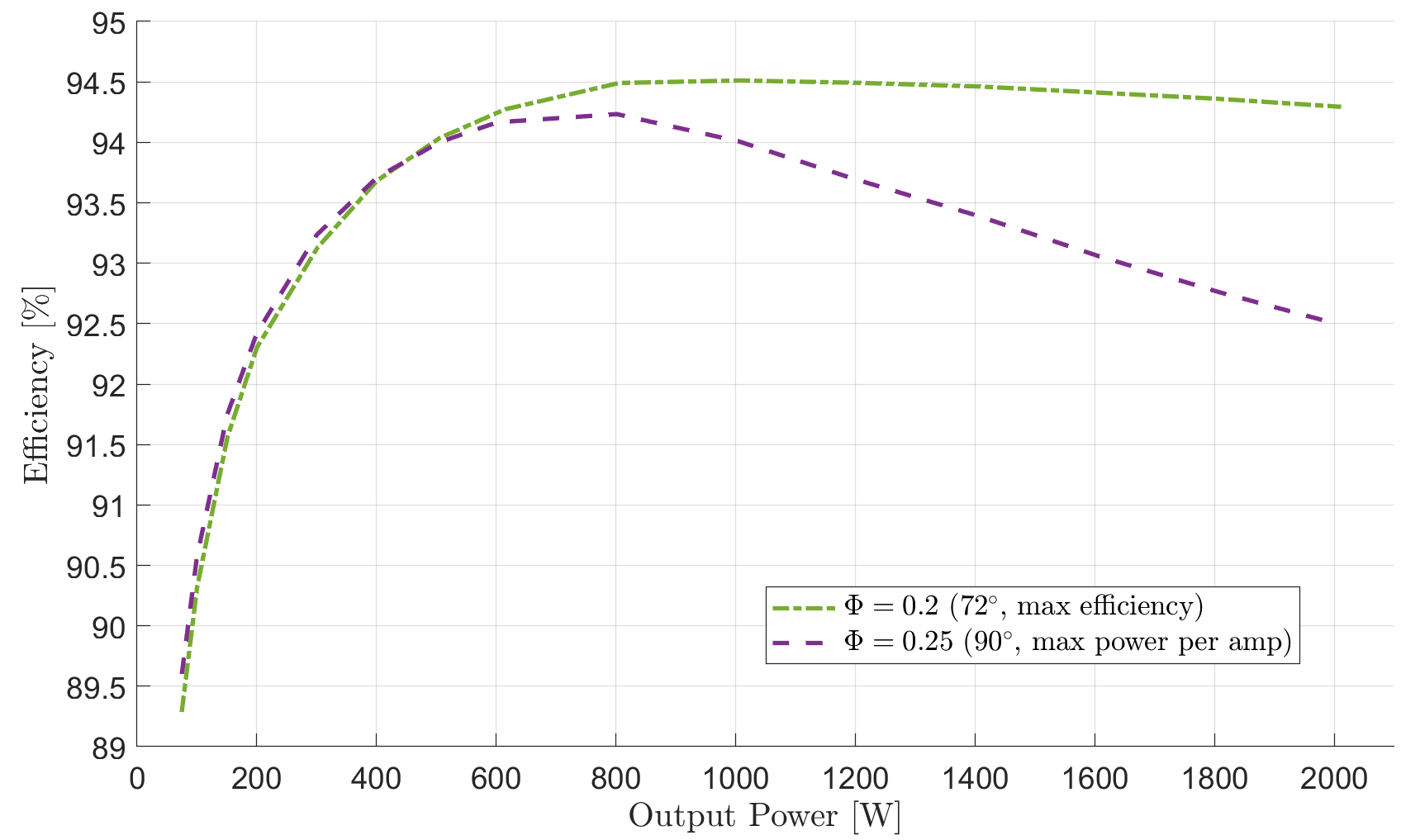}}
\caption{Prototype efficiency over output power range. Phase shift $\Phi$ for both maximum efficiency and maximum power per ampere.}
\label{fig:eff-vs-power}
\end{figure}

The analytical model curve and experimental data points correlate well, supporting the fidelity of the power flow analysis. The data show that the power flow peaks at a phase shift of 0.25 (90 degrees, marked with the purple dashed line), as expected. At this operating point, the measured DC-DC efficiency is 93.7\%. The maximum efficiency achieved was 94.3\%, at a phase shift of 0.20 (72 degrees, marked with the green dashed line). Note, the analytical model uses lossless components, thus we assume that overall efficiency will peak where the analytical model predicts the highest power factor, as that is where the most watts per volt-amperes are transferred and unnecessary $I^2R$ losses are minimized. The efficiency data points collected from experiments do follow this trend, and peak power and peak efficiency occurring at different phase shift angles prompted additional experiments where phase shift angle was fixed and primary DC current was increased, presented in Fig. \ref{fig:eff-vs-power}.

Fig. \ref{fig:eff-vs-power} plots measured DC-DC efficiency versus output power for the two fixed phase shift angles that correspond to maximum power transfer (90 degrees) and peak measured efficiency (72 degrees).  The maximum power angle curve (purple) efficiency peaks at 94.2\% with 5.3 A of primary DC link current and 800 W of output power. Beyond 800 W, efficiency quickly drops off in a linear fashion. The maximum efficiency angle curve (green) peaks at 94.5\% with 6.3 A of primary DC link current and 1 kW of output power. Interestingly, Fig. \ref{fig:eff-vs-power} also shows that when operating the prototype at the maximum efficiency angle, there is a much more gradual decline in efficiency as the converter moves into higher power operation. The operating point of 2 kW maintains a 94.3\% efficiency, with an input current of 9 A. By contrast, the efficiency for same power level at the 90 degree phase shift drops to 92.5\% though it only draws 8.5 A of input current for 2 kW of output power.

\begin{figure}[htbp]
\centerline{\includegraphics[width=0.95\columnwidth]{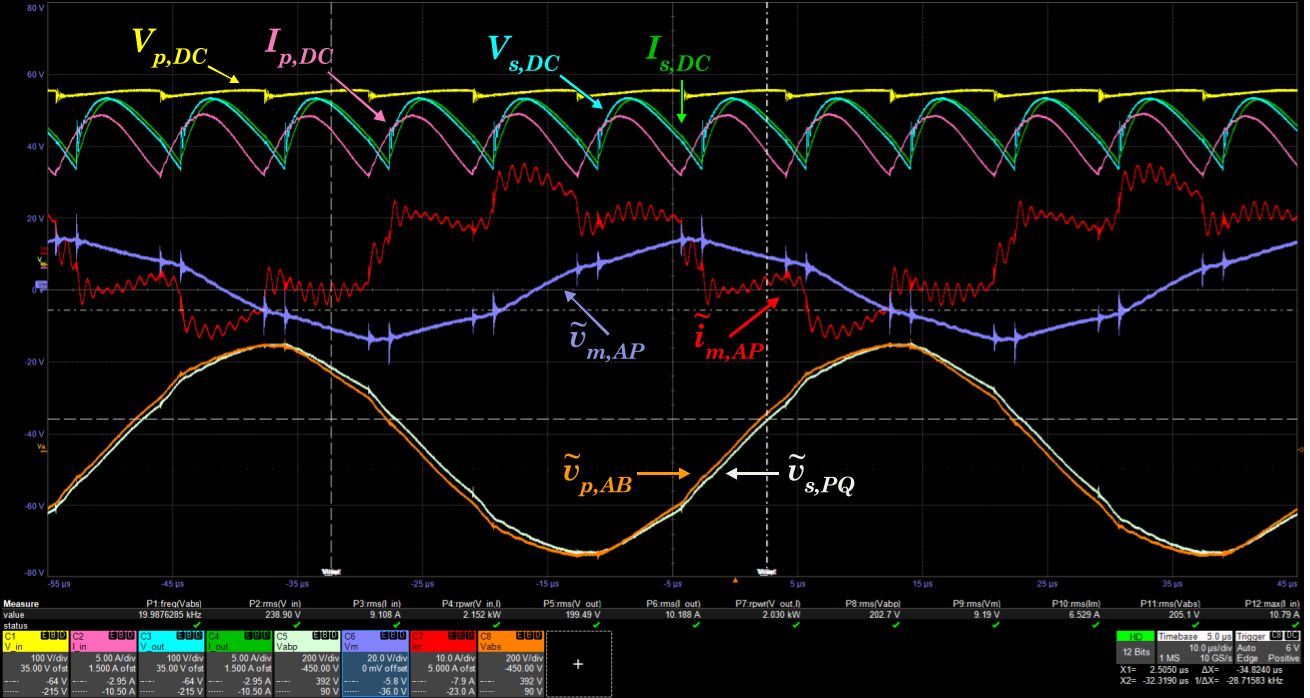}}
\captionsetup{font={color=black, footnotesize}, belowskip=-4pt}
\caption{Oscilloscope capture of SiC CBT prototype operating near unity gain and 2 kW of output power.}
\label{fig:scope-unity}
\end{figure}

Fig. \ref{fig:scope-unity} shows a time domain oscilloscope capture of the prototype operating near unity gain with 2 kW of output power and $V_{p,DC} = 240$ V. The shapes of the AC capacitor waveforms align with the underlying assumptions made for the analytical model in Section \ref{sec:model}. The ripple seen in the red mutual capacitor current is due to the parasitic inductance of the mutual capacitors.

\color{black}

\color{black}

\begin{figure}[htbp]
\captionsetup{font={color=black, footnotesize}, belowskip=-4pt}
\centerline{\includegraphics[width=0.9\columnwidth]{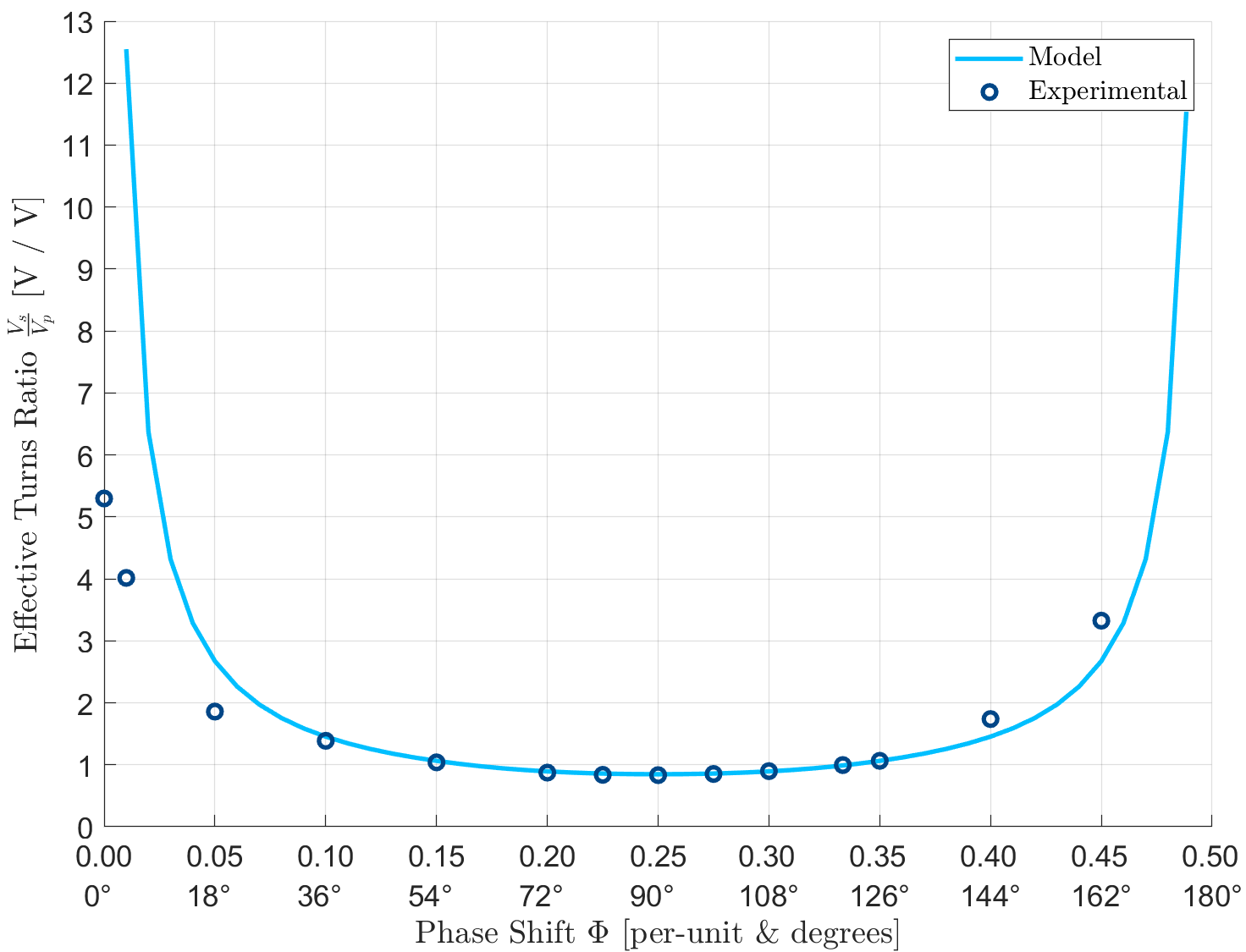}}
\caption{Analytical model prediction vs experimentally measured effective turns ratio.}
\label{fig:ratio-comparison}
\end{figure}

The measured and analytically modeled CBT DC-DC effective turns ratios \eqref{eq:A_ratio}, \eqref{eq:B_ratio}, \eqref{eq:C_ratio} versus primary-to-secondary phase shift are plotted in Fig. \ref{fig:ratio-comparison}. The measured and modeled values correlate well in the region from 36 - 162 degrees. Beyond that range, the experimental results diverge from the model given non-optimal parameters including saturation, ripple, and losses.

\begin{figure}[htbp]
\centerline{\includegraphics[width=0.9\columnwidth]{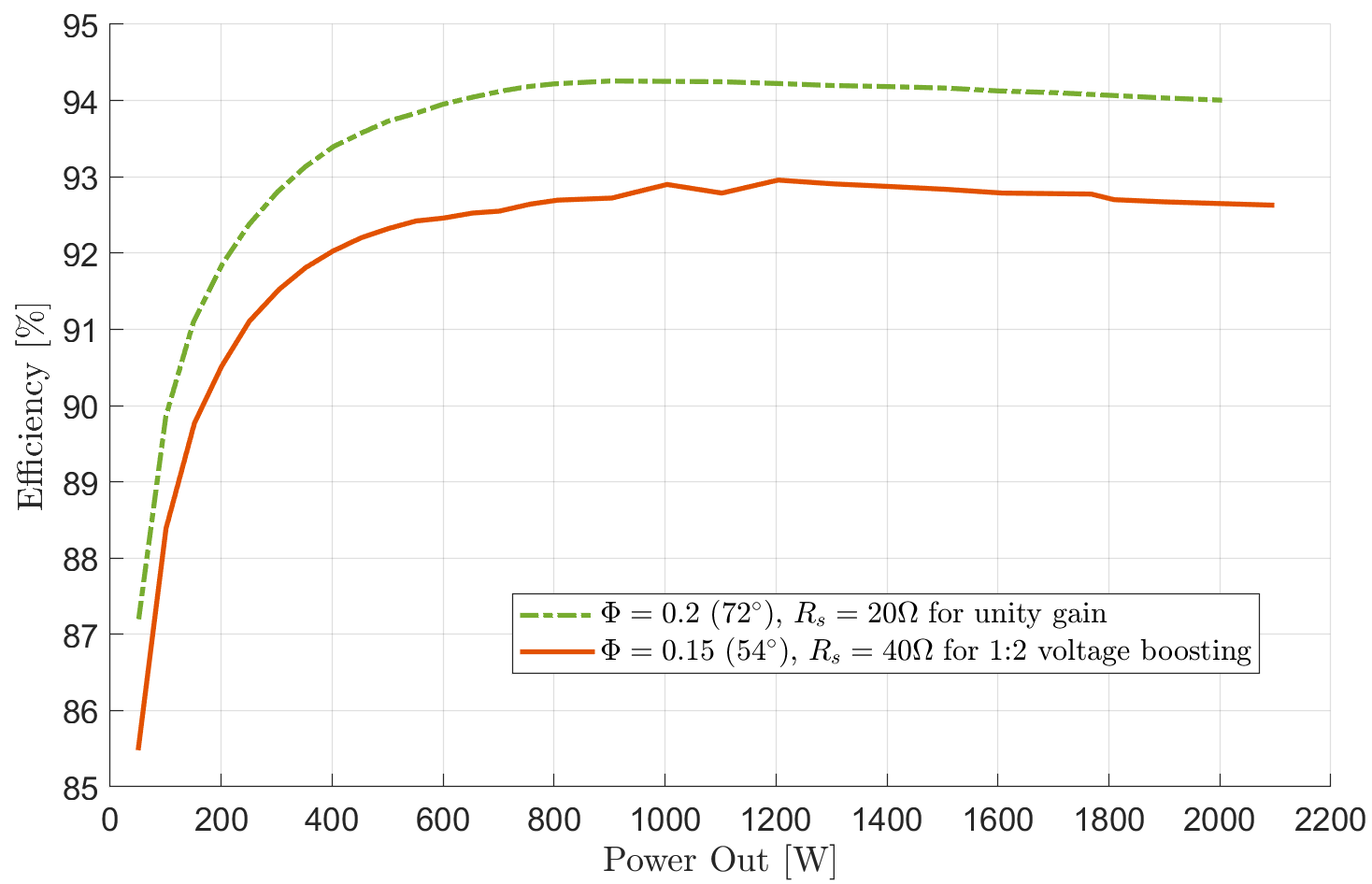}}
\captionsetup{font={color=black, footnotesize}, belowskip=-4pt}
\caption{Prototype efficiency vs output power, unity voltage gain versus 1:2 voltage boosting.}
\label{fig:efficiency-unity-vs-boost}
\end{figure}

\begin{figure}[htbp]
\centerline{\includegraphics[width=0.95\columnwidth]{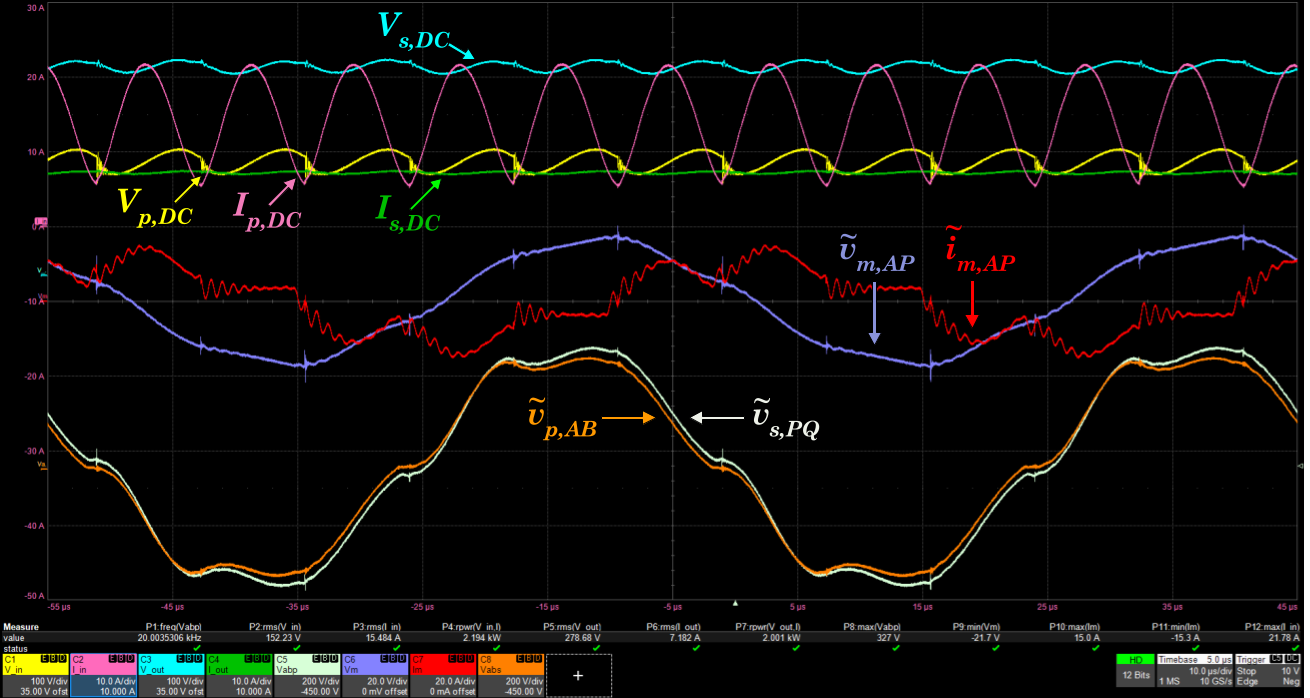}}
\captionsetup{font={color=black, footnotesize}, belowskip=-4pt}
\caption{Oscilloscope capture of SiC CBT prototype operating with 1:2 voltage gain and two kilowatts of output power.}
\label{fig:scope-boosting}
\end{figure}

As shown by the model, the maximum realistically achievable effective turns ratio of the CBT is somewhat limited. To demonstrate a voltage boost of approximately 1:2, the load resistance $R_s$ was increased from 20 $\Omega$ to 40 $\Omega$, and the phase shift was deviated to 54 degrees. The orange curve in Fig \ref{fig:efficiency-unity-vs-boost} shows how a limited efficiency penalty of 1\% was incurred over a wide power range. Fig \ref{fig:scope-boosting} shows a time domain oscilloscope capture of the prototype operating with a 1:2 voltage boost at 2 kW of output power and $V_{p,DC} = 150$ V.


\color{black}

While these experimental results verified the analytical model in its current form, they also point to areas that merit further development and investigation, including:

\begin{enumerate}
  \item Semiconductor losses – The peak efficiency could be increased substantially, by several percentage points, through use of bidirectional blocking monolithic devices to eliminate the series blocking diodes. \color{black} In the case of this prototype, the forward drop was around 1.3 V.
  \item \color{black} Reactive power reduction -- The internal reactive power circulating in the CBT is loss mechanism given the equivalent series resistance of the capacitors. Dual-phase-shift (DPS) control, adapted for the CBT from its presentation in \cite{Mi2008ReactivePower}, could be used to reduce or eliminate these losses. \color{black}
  \item Isolation \& fault limitations – implications due to common mode voltages and the grounding of the primary and secondary sides need to be established.
\end{enumerate}


\section{Conclusions}

An isolated dual active bridge with no AC magnetic components was developed by means of current source inverters and an AC capacitive bridge network, resulting in a circuit termed a capacitive bridge transformer (CBT). \color{black} An analytical model was developed for the three-phase CBT DC-DC power, internal apparent power, internal reactive power, and effective turns ratio. The equations show that the CBT exhibits classical DAB characteristics.

The CBT operational concepts and analytical model were confirmed with an experimental prototype operating at 2 kW and 94.3\% efficiency. The experimental data closely follows the trends predicted by the analytical model. Additional experiments were conducted to verify how altering phase shift or effective turns ratio affects the efficiency of this topology; notably, efficiency of this prototype is maximized at a phase shift of 72 degrees, and the converter achieved 92.6\% efficiency at 2 kW while boosting voltage 1:2.

A future prototype will make use of four-quadrant bidirectional switches with low on resistance to further increase efficiency by eliminating diode forward voltage drop. \color{black}

\section*{Acknowledgment}

The authors thank the Wisconsin Electric Machines and Power Electronics Consortium (WEMPEC) and its members for their continued support. 
\color{black}







\bibliographystyle{IEEEtran}
\bibliography{references-2}

\begin{thebibliography}{10}
\providecommand{\url}[1]{#1}
\csname url@samestyle\endcsname
\providecommand{\newblock}{\relax}
\providecommand{\bibinfo}[2]{#2}
\providecommand{\BIBentrySTDinterwordspacing}{\spaceskip=0pt\relax}
\providecommand{\BIBentryALTinterwordstretchfactor}{4}
\providecommand{\BIBentryALTinterwordspacing}{\spaceskip=\fontdimen2\font plus
\BIBentryALTinterwordstretchfactor\fontdimen3\font minus \fontdimen4\font\relax}
\providecommand{\BIBforeignlanguage}[2]{{%
\expandafter\ifx\csname l@#1\endcsname\relax
\typeout{** WARNING: IEEEtran.bst: No hyphenation pattern has been}%
\typeout{** loaded for the language `#1'. Using the pattern for}%
\typeout{** the default language instead.}%
\else
\language=\csname l@#1\endcsname
\fi
#2}}
\providecommand{\BIBdecl}{\relax}
\BIBdecl

\bibitem{Huber2016Solid-StateConcepts}
\BIBentryALTinterwordspacing
J.~E. Huber and J.~W. Kolar, ``{Solid-State Transformers: On the Origins and Evolution of Key Concepts},'' \emph{IEEE Industrial Electronics Magazine}, vol.~10, no.~3, 2016. [Online]. Available: \url{https://ieeexplore.ieee.org/document/7575768/}
\BIBentrySTDinterwordspacing

\bibitem{Doncker1991AApplications}
\BIBentryALTinterwordspacing
R.~D. Doncker, D.~Divan, and M.~Kheraluwala, ``{A three-phase soft-switched high-power-density DC/DC converter for high-power applications},'' \emph{IEEE Transactions on Industry Applications}, vol.~27, no.~1, 1991. [Online]. Available: \url{https://ieeexplore-ieee-org.ezproxy.library.wisc.edu/document/67533/}
\BIBentrySTDinterwordspacing

\bibitem{Guillod2020Medium-frequencyAnalysis}
T.~Guillod and J.~W. Kolar, ``{Medium-frequency transformer scaling laws: Derivation, verification, and critical analysis},'' \emph{CPSS Transactions on Power Electronics and Applications}, vol.~5, no.~1, pp. 18--33, 3 2020.

\bibitem{Leibl2017DesignApplications}
M.~Leibl, G.~Ortiz, and J.~W. Kolar, ``{Design and Experimental Analysis of a Medium-Frequency Transformer for Solid-State Transformer Applications},'' \emph{IEEE Journal of Emerging and Selected Topics in Power Electronics}, vol.~5, no.~1, pp. 110--123, 3 2017.

\bibitem{Soltau2014IronConverter}
N.~Soltau, D.~Eggers, K.~Hameyer, and R.~W. De~Doncker, ``{Iron losses in a medium-frequency transformer operated in a high-power DC-DC converter},'' \emph{IEEE Transactions on Magnetics}, vol.~50, no.~2, pp. 953--956, 2014.

\bibitem{Yin2021ImpactConverter}
S.~Yin, S.~Debnath, R.~Wojda, P.~Marthi, and M.~Saeedifard, ``{Impact of the Transformer Magnetizing Inductance on the Performance of the Dual-Active Bridge Converter},'' \emph{2021 IEEE 22nd Workshop on Control and Modelling of Power Electronics, COMPEL 2021}, 2021.

\bibitem{Nowakowski2025CBT}
P.~Nowakowski, M.~Gupta, and D.~C. Ludois, ``Capacitive bridge transformer: An isolated, current-stiff dual active bridge architecture,'' in \emph{2025 IEEE Energy Conversion Conference Congress and Exposition (ECCE)}, 2025, pp. 1--8.

\bibitem{Brooks2024OnDesigns}
N.~C. Brooks, J.~Zou, S.~Coday, T.~Ge, N.~M. Ellis, and R.~C.~N. Pilawa-Podgurski, ``{On the Size and Weight of Passive Components: Scaling Trends for High-Density Power Converter Designs},'' \emph{IEEE Transactions on Power Electronics}, vol.~39, no.~7, pp. 8459--8477, 7 2024.

\bibitem{Peterson1978US4079305ALoads}
\BIBentryALTinterwordspacing
H.~A. Peterson and N.~Mohan, ``{US4079305A - Power supply for high power loads},'' 3 1978. [Online]. Available: \url{https://patents.google.com/patent/US4079305A/en}
\BIBentrySTDinterwordspacing

\bibitem{Kustom1980ThyristorCoils}
R.~L. Kustom, \emph{{Thyristor Networks for the Transfer of Energy between Superconducting Coils}}.\hskip 1em plus 0.5em minus 0.4em\relax Madison, WI: University of Wisconsin Press, Ltd., 1980.

\bibitem{Fuja1980Three-PhaseCoils}
R.~E. Fuja, R.~L. Kustom, and M.~Ehsani, ``{Three-Phase Energy Transfer Circuit with Superconducting Energy Storage Coils},'' \emph{IEEE Transactions on Industry Applications}, vol. IA-16, no.~3, pp. 438--444, 1980.

\bibitem{Ehsani1985AMagnets}
M.~Ehsani, R.~L. Kustom, and R.~W. Boom, ``{A one-phase dual converter for two quadrant power control of superconducting magnets},'' \emph{IEEE Transactions on Magnetics}, vol.~21, no.~2, pp. 1115--1118, 1985.

\bibitem{Ehsani1993InverseApplications}
M.~Ehsani, I.~Husain, and M.~O. BilgiC, ``{Inverse Dual Converter (IDC) for High-Power DC-DC Applications},'' \emph{IEEE Transactions on Power Electronics}, vol.~8, no.~2, pp. 216--223, 1993.

\bibitem{Bilgic2003AnalysisStorage}
M.~Bilgic and M.~Ehsani, ``{Analysis of inductor-converter bridge by means of state-space averaging technique (superconducting magnet energy storage)},'' pp. 116--121, 1 2003.

\bibitem{Baliga2023TheConverters}
B.~J. Baliga, D.~Hopkins, S.~Bhattacharya, A.~Agarwal, T.~H. Cheng, R.~Narwal, A.~Kanale, S.~S. Shah, and K.~Han, ``{The BiDFET Device and Its Impact on Converters},'' \emph{IEEE Power Electronics Magazine}, vol.~10, no.~1, pp. 20--27, 3 2023.

\bibitem{Zhang2016AApplication}
H.~Zhang, F.~Lu, H.~Hofmann, W.~Liu, and C.~C. Mi, ``{A Four-Plate Compact Capacitive Coupler Design and LCL-Compensated Topology for Capacitive Power Transfer in Electric Vehicle Charging Application},'' \emph{IEEE Transactions on Power Electronics}, vol.~31, no.~12, pp. 8541--8551, 12 2016.

\bibitem{LudoisDai2015Survey}
J.~Dai and D.~C. Ludois, ``A survey of wireless power transfer and a critical comparison of inductive and capacitive coupling for small gap applications,'' \emph{IEEE Transactions on Power Electronics}, vol.~30, no.~11, pp. 6017--6029, 2015.

\bibitem{Dai2015SingleTransfer}
------, ``{Single active switch power electronics for kilowatt scale capacitive power transfer},'' \emph{IEEE Journal of Emerging and Selected Topics in Power Electronics}, vol.~3, no.~1, pp. 315--323, 2015.

\bibitem{Kamalapur2025DimensionalConverters}
A.~Kamalapur, J.~S. Bae, M.~Cairnie, R.~Nassar, J.~Knoll, D.~Boroyevich, G.~Q. Lu, C.~Dimarino, Q.~Li, and K.~D. Ngo, ``{Dimensional Parasitics Absorption in Capacitively-Isolated {\'{C}}uk Converter for Medium-Voltage High Step-Down Converters},'' \emph{Conference Proceedings - IEEE Applied Power Electronics Conference and Exposition - APEC}, pp. 2228--2235, 2025.

\bibitem{Channegowda2020HighCharacteristics}
P.~Channegowda, P.~Rengasamy, D.~Ludois, and G.~Venkataramanan, ``{High Voltage DC- DC Converter Topology based on Capacitive Coupling with Soft Switching Characteristics},'' \emph{9th IEEE International Conference on Power Electronics, Drives and Energy Systems, PEDES 2020}, 12 2020.

\bibitem{Neuner2023AnalysisTransformer}
D.~Neuner, M.~Hartmann, J.~W. Kolar, and J.~Huber, ``{Analysis of a Solid-State Transformer Employing Capacitively Isolated Series-Stacked Converter Cells and a Single Medium-Frequency Transformer},'' \emph{2023 IEEE 24th Workshop on Control and Modeling for Power Electronics, COMPEL 2023}, 2023.

\bibitem{Mi2008ReactivePower}
H.~Bai and C.~Mi, ``Eliminate reactive power and increase system efficiency of isolated bidirectional dual-active-bridge dc–dc converters using novel dual-phase-shift control,'' \emph{IEEE Transactions on Power Electronics}, vol.~23, no.~6, pp. 2905--2914, 2008.

\end{thebibliography}

\end{document}